\documentclass{aa}  
\usepackage{graphicx}
 \usepackage[latin1]{inputenc}
 \usepackage{natbib}
\bibpunct{(}{)}{;}{a}{}{,} % to follow the A&A style
\def\kms{~km~s$^{-1}$}
\def\kmskpc{~km~s$^{-1}$~kpc$^{-1}$}

\def\deg{^\circ}
\def\lvd{$lv$-diagram}
\def\Msun{M$_\odot$}

\begin{document}
   \title{Gas flow models in the Milky Way embedded bars}

%   \subtitle{I}

   \author{N. J.  Rodriguez-Fernandez  \inst{1}   
          \and F. Combes \inst{2}
     }

   \offprints{N. J. Rodriguez-Fernandez}

   \institute{IRAM, 300 rue de la Piscine, 38406 St. Martin d'Heres, France \\
    \email{rodriguez@iram.fr}
    \and LERMA,  Observatoire de Paris,  61 Av de  l'Observatoire, 75014 Paris, France
      }

   \date{Received September 15, 1996; accepted March 16, 1997}

% \abstract{}{}{}{}{} 
% 5 {} token are mandatory
 
%\authorrunning{Rodriguez-Fernandez }
%\titlerunning{Dynamics and molecular chemistry in the Galactic center}

  \abstract
   { The gas distribution and dynamics in the inner Galaxy present many unknowns as the origin of the asymmetry of the \lvd ~ of the Central Molecular Zone  (CMZ). On the other hand, there are recent evidences in the stellar component of the presence of a nuclear bar  that could be slightly lopsided. }
   {Our goal is to characterize the nuclear  bar observed in 2MASS maps and to study the gas dynamics in the inner Milky Way  taking into account this secondary bar.}
   { We have derived a realistic mass distribution by fitting the 2MASS star counts map of Alard (2001) with a model including three  components (disk, bulge and nuclear bar) and we have simulated the gas dynamics,  in the deduced gravitational potential, using a sticky-particles code.}
   { Our simulations of the gas dynamics reproduce successfully 
the main characteristics of the Milky Way for a bulge orientation  of $20\deg-35\deg$ with respect to the Sun-Galactic Center (GC) line and a pattern speed of 30-40 \kmskpc. In our models the Galactic Molecular Ring (GMR) is not an actual ring but the inner parts of the spiral arms, {while the 3-kpc arm and its far side counterpart are lateral arms that contour the bar.}
Our simulations reproduce, for the first time, the parallelogram shape of the \lvd ~ of the CMZ as the gas response to the nuclear bar. This bar  should be oriented by an angle of $\sim 60\deg-75\deg$ with respect to the Sun-GC line and its mass amounts to $(2-5.5)\,10^9$ \Msun.
We show that the observed asymmetry of the CMZ cannot be due to  lopsidedness of the nuclear bar as suggested by the 2MASS maps. }
{We do not find clear evidences of lopsidedness in the stellar potential. We propose that the observed asymmetry of the central gas layer can be due to the
infalling of gas  into the CMZ in the {\it l=1.3$\deg$-complex}.}

\keywords{Galaxy: structure -- Galaxy: center -- Galaxy: kinematics and dynamics -- ISM: kinematics and dynamics -- Methods: numerical  }

\maketitle

%
%%%%%%%%%%%%%%%%%%%%%%%%%%%%%%%%%%%%%%%%%

\section{Introduction}

Studying the structure and the dynamics of the Milky Way galaxy is necessary to understand its formation and evolution. For example, the large scale star formation rate, the gas transfer to the nucleus and the fueling of the central super-massive black hole are strongly influenced by the large scale dynamics of the Galaxy.

At least three components are needed to explain the rotation curve of the Galaxy: a disc, a bulge and a massive dark halo.
In near-infrared images, the bulge is a triaxial structure difficult to distinguish from a bar.
The first direct evidence of the barred nature of the Milky Way was found by \cite{Blitz91}.
The analysis of COBE/DIRBE data \citep{Dwek95, Binney97, Freudenreich98} supported Blitz \& Spergel results.
\cite{Dwek95} found that the peanut-shaped bulge seen in the COBE image is best fitted by {\it boxy} Gaussians functions.
Since COBE, the direct evidence of the bar has increased based on photometric surveys and star counts in the near-infrared \citep[see for instance][and references therein]{Lopez-Corredoira00, Lopez-Corredoira05, Babusiaux05}.
Although, there is a consensus on the triaxial nature of the bulge, the actual shape and the inclination  is still under debate with currently no clear agreement among different authors.
The inclination angles found range from $\sim 10\deg-15\deg$ \citep{Binney91, Freudenreich98, Lopez-Corredoira97} to $\sim 30\deg-45\deg$ \citep{Weiner99, Sevenster99, Lopez-Corredoira01} with a number of works giving values in the range $15\deg-30\deg$ \citep{Mulder86, Binney97, Fux99, Englmaier99, Bissantz02, Bissantz03, Babusiaux05, Lopez-Corredoira05}.

In addition to the bar-like {\it boxy}-bulge, some recent works have also proposed the existence of another bar that would be longer and thinner than the bulge \citep{Picaud03,  Lopez-Corredoira01, Lopez-Corredoira07,  Benjamin05}. As the bulge, the near side of this long and thin bar will also be located in the first Galactic quadrant. In contrast, the long bar would be oriented at an angle of $\sim 45\deg$ with respect to the line of sight towards Galactic center.
In the present paper we will not discuss this component as the mass should be dominated by the bulge.
Hereafter, we will refer to the triaxial boxy bulge as ``large bar" or simply ``bulge".

Due to the difficulty to measure the stellar velocities, most of our knowledge of the Galactic dynamics come from spectroscopic observations of the atomic and molecular interstellar medium \citep{Rougoor60, Scoville72, Liszt78, Bally87, Dame01}.
The gas exhibit large non-circular motions in the inner Galaxy. Among the possible explanations for these kinematics, de Vaucouleurs (1964) proposed the gas response to a bar potential. This explanation has also been discussed in other observational works as those of \cite{Kerr67}, \cite{Bania77} or \cite{Liszt78}.
Nevertheless, it is the comparison of the observations with numerical simulations of the gas dynamics what have allowed to better understand the structure of the Galaxy and
to derive important dynamical information as the pattern speed of the bar.
The first attempt was that of \cite{Mulder86} solving the gas-dynamical equations to get the gas flow in a weak bar for the whole Galaxy.

{
A major step in our understanding of the gas kinematics in the Galactic Center was
the paper by \cite{Binney91}.
These authors made a comparison of the CO(1-0) and HI longitude-velocity  diagrams (hereafter \lvd) of \cite{Bally87} and \cite{Burton78}, with the $lv$ locus of closed stellar orbits in a bar potential (x1 and x2 orbits).
The \lvd \  of the molecular gas in the inner $\sim 4 \deg$ of the Galaxy shows a parallelogram shape that resembles the $lv$ locus of the \textit{cusped} x1 orbit for a bar with inclination of $\sim 16 \deg$.
In spite of the globally successful interpretation of the CO and HI \lvd s, the \cite{Binney91} model cannot account for the observed asymmetry of the CO  parallelogram, which is lopsided towards positive longitudes while the modeled parallelograms are centered at longitude $0\deg$.
In addition the expected upper right vertex of the locus of the \textit{cusped} x1 orbit has a velocity close to 0, while the upper right vertex of the observed CO parallelogram shown in Fig. 2 of \cite{Binney91} has velocity of $\sim 150$ \kms. }

The qualitative insight on the gas dynamics that one could obtain by comparing the \lvd~ of the gas with that of a few stellar orbits, should be tested performing simulations of the gas dynamics that take into account the dissipative nature of the gas clouds. This was done by \cite{Jenkins94}, who tried to reproduce the \lvd ~ of the CMZ using a sticky-particles code but with moderate success.
{ Using a combination of SPH and collisionless particles, \cite{Lee99} were able to reproduce an inner ring with the right inclination in the \lvd \ and also some of the gas clouds with non-circular velocities in the range $l<|6\deg|$.} 

To our knowledge, the works by \cite{Binney91}, \cite{Jenkins94} and \cite{Lee99} are the only dynamical models that have tried to interpret the dynamics of the complex Galactic center region (the central 500 pc).
Other models of the Galactic dynamics have been mainly devoted to larger scales, from the works of \cite{Mulder86} and Wada et al (1994) to the more recent works by \cite{Fux99}, \cite{Englmaier99}, \cite{Weiner99} and \cite{Bissantz03}. The bar pattern rotation speed for most of those models are in the range from 40 to 60 \kmskpc.  Regarding the corotation radius, 
it is estimated to be in the range 3-4 kpc by some works \citep{Englmaier99, Bissantz03, Habing06} and in the range 4-5 kpc by other works \citep{Combes96, Sevenster99, Fux99}.
In contrast, \cite{Binney91} placed the corotation at 2.4 kpc and found a pattern speed of 80 \kmskpc.
As pointed out by \cite{Combes96}, the discrepancy between Binney et al. estimation and other  estimations can be due to the fact that the molecular gas distribution in the Galactic center region used by Binney et al. could be the response to a small nuclear bar instead of the large  bar of the Galaxy.

Indeed, there are recent evidences of the presence of a distinct structure in the nuclear region of the Galaxy. The star counts map made by \cite{Alard01} with 2MASS data shows an excess of counts in the inner four degrees of the Galaxy that resembles a small nuclear bar. This bar could be lopsided with the barycenter displaced to negative longitudes.
\cite{Nishiyama05} have also found evidence of a structure different to the large bar or bulge in the inner degrees of the Galaxy.  

In the present paper, we want to investigate the gas response to the mass distribution inferred from the 2MASS data. In particular, we model the central mass distribution as a small nuclear bar. 
We are interested in the role of this nuclear bar to explain the differences among the results of previous works \citep{Binney91, Fux99, Combes96}.
We also want to study whether a lopsided nuclear bar could explain the apparent lopsidedness of the molecular gas distribution in the nuclear region, whose complex dynamics are far from being understood.
Therefore, we have modeled the 2MASS star counts maps of Alard fitting different parametric functions using three components: disc, bulge/bar and nuclear bar.
We use that mass distribution to determine the galactic stellar potential and to  model the gas dynamics using a N-body code that takes into account cloud collisions.

{ Nested bars are common in external galaxies. As much as one third of barred galaxies have a secondary nuclear bar \citep{Laine02}. In some cases the secondary bar is coupled with the primary bar \citep{Shaw93} while in other cases the two bars rotate with different speeds \citep[][]{Wozniak95}.
The present paper is the first one dealing with the gas flow in the Milky Way nested bars and we will assume that both bars are dynamically coupled (they rotate with the same speed). 
As we discuss in Sect. 8, this seems a good assumption since our simulations explain for the first time some characteristics of the inner Galaxy.
}

The paper is organized as follows: in the next section we summarize our understanding of the distribution and kinematics of the interstellar gas. 
In Sects. 3 and 4 we present the 2MASS star counts model and the fits to the data.
The gas dynamics simulations are described in Sect. 5, and the results are presented in Sects. 6 an 7. We discuss the nature of the nuclear bar in Sect. 8. Our conclusions are summarized in Sect. 9.

\begin{figure*}[ht!]
\centerline{\includegraphics[angle=-90,width=16cm]{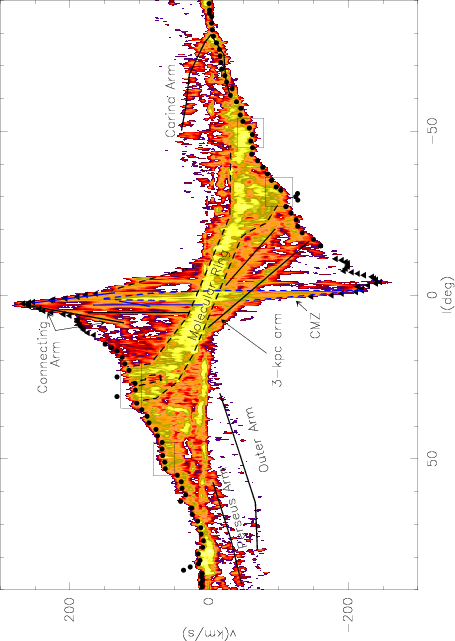}}
\caption{Longitude-velocity ($lv$) diagram of the CO(1-0) emission \citep{Dame01}. The solid lines trace the position of some remarkable features as the locus of the spiral arms, the 3-kpc arm and the Connecting Arm. {  The black dashed lines indicate the contour of the Galactic Molecular Ring. The solid circles are the terminal velocities measurements of \cite{Fich89} using CO while the triangles are the terminal velocities determined from the HI data of \cite{Burton93}. 
The boxes mark the position of the Sagittarius, Scutum, Norma and Centaurus tangent points, located respectively at $l \sim 50 \deg ,\ 30 \deg, \, -30\deg$ and $-50\deg$.
The lines concerning the Nuclear Disk or Central Molecular Zone are blue but they are better shown in the next figure. } All these lines can be used to compare with the figures showing the simulations results. }
\label{fig:dame}
\end{figure*}

%%%%%%%%%%%%%%%%%%%%%%%%%%%%%%%%%%%%%%%%%%%5
\section{The distribution and kinematics of the interstellar gas}
In this section we summarize outstanding structures in the  longitude-velocity diagram (hereafter $lv$-diagram) of the interstellar gas.
The large scale $lv$-diagram derived from observations of the CO(1-0) line by \citep{Dame01}
is shown in Fig. \ref{fig:dame} while Fig. \ref{fig:bally} shows the $lv$-diagram of the nuclear region derived also from CO(1-0) observations by \cite{Bally87}.
We also discuss the plausible face on view of the Galaxy. The different features discussed here will be used in the following sections to compare with our simulations of the gas dynamics.

{
\subsection{The terminal velocity curve}
The terminal velocity curve (TVC) is not a kinematic feature in itself as many independent features contribute to this curve at different Galactic longitudes.
However, we discuss it here since it has been used to constrain several dynamical models of the Milky Way \citep{Binney91,Weiner99, Englmaier99, Bissantz03}.
There has been different determinations of the TVC using CO and HI data \citep{Fich89, Burton93, Clemens85, Alvarez90}.
The dispersion on the TVC points derived by different authors is typically 10 \kms ~ \citep[see for instance Fig. 6 of][]{Bissantz03}.  
The TVC measurements by \cite{Fich89}  and \cite{Burton93} are shown in Fig. \ref{fig:dame} and in the figures showing the simulations results.
In addition, in Fig. \ref{fig:dame} we have indicated with boxes  the Sagittarius, Scutum, Norma and Centaurus tangent points, located respectively at $l \sim 50 \deg ,\ 30 \deg, \, -30\deg$ and $-50\deg$ \cite[see for instance][]{Dame01, Englmaier99}.  
}

\subsection{Spiral arms}
Many works have studied the spiral structure of the Galaxy using different tracers as HII regions, atomic and molecular gas or photometric observations in different bands from the near to the far infrared.
The different models of the spiral arms have two (\cite{Bash81} using HII regions,  \cite{Drimmel01} using K-band observations, \cite{Benjamin08} using Spitzer), three \citep[][using HI]{Nakanishi03, Kulkarni82} or four arms (\cite{Georgelin76} using HI regions;  \cite{Nakanishi06} using CO or  \cite{Drimmel00} using the 240 microns luminosity). 
Deprojecting a given data set to find out the face on view of the Galaxy is a difficult task in particular for the inner Galaxy. For instance, the kinematical distances based in the rotation curve are not accurate due to the presence of important non-circular motions.
Furthermore, after performing the deprojection there is an uncertainty in how different structures are associated to a common arm \citep[see][]{Nakanishi06}. Different authors have even fitted different spirals to the same data. 
A plausible explanation for the discrepancy among the two arms derived from stellar tracers (K band) and the four arms that are found in most of the interstellar tracers is that the Milky Way has two stellar spiral arms but the gas response  is a four-armed spiral \citep{Drimmel00}.
In any case, the major arms are  logarithmic with pitch angles of $10\deg-15\deg$. In addition, to these grand-design spirals there is a more 
flocculent structure giving a number of minor arms like the Local Arm, which is a small arm linked to the Perseus spiral arm.
In order to compare with our simulations, in Figs. \ref{fig:simu_a70}, \ref{fig:simu_o30} and \ref{fig:simu_a70_sche} we have traced the sketch of the logarithmic spiral arms of  \cite{Nakanishi06}, which are very similar to the classical spirals of \cite{Georgelin76}.
The locus of different spiral arms in the $lv$-diagram are shown in Fig. \ref{fig:dame}. The Carina arms deserves a special mention since the Carina tangent point is clearly seen at $l \sim -80\deg$.

{ The determinations of the  speed of the spiral pattern give values in the range from 13.5 to 59 \kmskpc \ \citep{Lin69, Fernandez01, Bissantz03, Martos04, Debattista02}. Thus, the spiral pattern speed could be lower that the bar  one. }

\subsection{The Galactic Molecular ring}
The so-called Galactic Molecular Ring (GMR) is one of the most prominent structures in the lv-diagram (Fig. \ref{fig:dame}). The locus of the GMR is more or less a diagonal line that passes trough the origin as a circular ring would do.
If the GMR is a real ring, it could be associated to a resonance. 
\cite{Binney91} proposed that the GMR could be located at the  Outer Linblad Resonance.
Other models that place the corotation farther out have suggested that the GMR could be better explained by the Ultra Harmonic Resonance \citep{Combes96}.

However, the real spatial distribution of the GMR is not known and it could be composed by imbricated spiral arms instead of being an actual ring.
Some authors \citep[see for instance][]{Englmaier99} consider that, if formed by several arms, the GMR should be located outside of the corotation since they exhibit almost circular velocities.
In their face-on deprojection of the neutral and molecular gas distribution, \cite{Nakanishi06} found that the GMR seems to be the inner part of the Sgr-Carina arm (the far-side  of the GMR) and the Scutum-Crux arm (the near-side of the GMR).

\subsection{Inner arms: the 3-kpc arm}
The 3-kpc arm is clearly seen in the $lv$-diagram of Fig. \ref{fig:dame} with a velocity of -53 km\,s$^{-1}$ at $l=0\deg$.
The name of this  feature comes from a tangent point at $l=-22\deg$ that 
corresponds to a distance of $\sim 3$ kpc. 
This arm is located in between the Sun and the GC since the clouds in the arm absorbs the radiation from the continuum sources in the GC.  
The 3-kpc arm cannot be explained with a combination of rotation and expansion \citep{Burke64} and has been interpreted as a stationary density wave in a barred potential \citep{Mulder86}. The simulations of \cite{Fux99} suggest that it is a lateral arm that surrounds the bar while, in the simulations by \cite{Englmaier99}, it would be an small arm arising from the extremity of the bar.

\cite{Sevenster99} and \cite{Habing06} propose a different explanation.
\cite{Sevenster99} found old stars associated with the 3-kpc arm. 
They interpret the fact that both old stars and gas follow the same trajectories as the probe
that the arm would be the locus of closed orbits and not a spiral density wave maximum. They propose that the 3-kpc arm has its origin near one of the two points where the bar meets its corotation radius and that the arm can be a channel to transport gas from the corotation to the GC and to fuel the star formation in cloud complexes like  Sgr B2. 

{ The recent discovery of the far-side counterpart of the 3-kpc arm by \cite{Dame08b} will certainly contribute to understand the exact nature of these features (see Sect. 6.3).}

\subsection{The connecting arm and Bania's clumps : tracing the dust lanes ? }
The ``Connecting Arm" is clearly seen in Fig. \ref{fig:dame} as a feature with high inclination with a velocity of $\sim$ 100 km\,s$^{-1}$ at $l=10\deg$ and more than 250 km\,s$^{-1}$ at $l=3\deg$ .
The name ``Connecting Arm" comes from the fact that it seems to be connected to the nuclear ring.
This a very puzzling structure that has not been discussed much in the literature \citep[see][for a short summary]{Fux99}.
On the one hand, \cite{Burton78} and \cite{Liszt80} suggested that the Connecting arm could be the edge of the HI ring found at a radius of $\sim 1$ kpc (see below). 
On the other hand, in one of his simulations, \cite{Fux99} obtained a feature in the $lv$-diagram  that resembles the Connecting Arm.
In that simulation, the Connecting Arm-like feature is the locus of the off-axis shocks or dustlanes (since in the optical images of barred galaxies the off-axis shocks are seen as dust lanes).
\cite{Marshall08} also interpret the Connecting Arm as the near side dustlane. 

\begin{figure}[th!]
\centerline{\includegraphics[width=8cm]{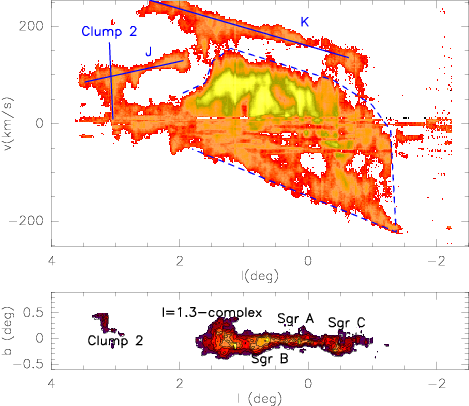}}
\caption{{\it Upper panel:} Longitude-velocity ($lv$) diagram of the CO(1-0) emission in the Central Molecular Zone using data of \cite{Bally87} and integrating all negative Galactic latitudes. The solid lines trace the position of some remarkable features as the Clump 2, or structures  K and J of \cite{Rodriguez-Fernandez06}. The dashed line indicates the contour of the very crowded central region.
{\it Lower panel:} Integrated intensity map.}
\label{fig:bally}
\end{figure}

However, in other of the simulations discussed by \cite{Fux99} the locus of the dustlane in the $lv$-diagram is a vertical feature extending approximately from a velocity of 0 km\,s$^{-1}$ to 200 km\,s$^{-1}$ at an almost constant longitude (see his figures 15 and 16). 
This reminds the ``clumps" discussed by \cite{Bania77}. These clumps  are outstanding cloud complexes found at $l=5.5\deg$ and $l=3.1\deg$  that exhibit a huge velocity dispersion of $\sim 200$ km\,s$^{-1}$ in a very narrow range of longitude.
A portion of the clump located at $l=3.1\deg$ (known as Clump 2) is showed in Fig. \ref{fig:bally}.
The Clump 2 has been interpreted as a dustlane by \cite{Stark86}, while in the \cite{Fux99} interpretation, Banias's clumps would be gas clouds that are about to enter the dustlanes shocks, although the actual dustlane would be the Connecting Arm as mentioned above. 
{ This is also the interpretation of \cite{Liszt06, Liszt08}, who have studied in detail the spatial and velocity structure of all these features.}

\subsection{The HI ring}
The HI emission  in the GC region can be modeled as a tilted circular ring with inner radius of $\sim$ 300 pc and outer radius of 1-1.5 kpc \citep{Burton78}. This first model was improved by \cite{Liszt80}, who discussed that an elongated ring could also explain  the observations.
The spatial distribution of the interstellar gas in the innermost 3 kpc of the Galaxy has recently been reviewed by \cite{Ferriere07}.
{ The model that they consider more plausible, despite its inherent uncertainties, is that of \cite{Liszt80} since an elongated ring seems to be easier to understand if it is composed by gas clouds moving in elongated x1 orbits.
Therefore, they modeled the HI ring as an elliptical ring with semi-major axis of 1.6 kpc, and axis ratio of 3.1, with an inner hole of semi-axis 800 pc $\times$ 258 pc that encloses the Central Molecular Zone (see bellow).
}

\subsection{The Central Molecular Zone}

The Central Molecular Zone (hereafter CMZ) refers to the central accumulation of gas in the inner hundreds of parsec of the Galaxy. 
This gas is mainly molecular and extends continuously approximately from $-1.5\deg$ to $2\deg$ (see Fig.~\ref{fig:bally}). The spatial distribution is not symmetric due to the prominent cloud complex located at $l=1.3\deg$  (hereafter {\it l=1.3$\deg$-complex}) that lacks of a negative longitude counterpart.
Figure~\ref{fig:bally} also shows the $lv$-diagram of the CMZ. This plot has been obtained integrating the CO(1-0) data of \cite{Bally87} for negative Galactic longitudes to remark a number of kinematical component as the Clump~2 (at $l=3.1\deg$), and the features K and J discussed by \cite{Rodriguez-Fernandez06}.

The kinematics of the inner region are very complex with a high fraction of the gas exhibiting non-circular velocities.
Figure \ref{fig:bally} shows with a dashed line the contour of the inner CMZ that contains the feature M of \cite{Rodriguez-Fernandez06} and the inner arms found by \cite{Sofue95}.
This contour resembles a parallelogram but it is not the parallelogram that inspired the \cite{Binney91} work.

The whole \lvd ~ shown in Fig. \ref{fig:bally} is very different to that used by \cite{Binney91}. 
{
These authors integrated the CO data of \cite{Bally87} over positive and negative latitudes and studied the region $2.2\deg >l > -2.2\deg$.
In their Fig. 2, the feature K is not clearly detached from the inner $lv$ parallelogram and they considered that it is part of a larger $lv$ parallelogram.
However, following the analysis of the \cite{Bally87} data by  \cite{Rodriguez-Fernandez06}, we reckon that  feature K is a distinct kinematic feature. 
This conclusion is also supported by the  large scale ($13\deg > l > -12\deg$) CO(1-0) survey  by \cite{Bitran97}, whose \lvd ~ shows clearly that the feature K extends  to $l \sim 3\deg$ where it is connected to the Connecting Arm.

Interestingly, when one compares the $lv$ parallelograms of the x1 orbits  displayed in Fig.~3 of \cite{Binney91} to the parallelogram shown in our Fig.~\ref{fig:bally}, the upper right vertex are in good agreement, showing velocities of $\sim 0$ \kms. 
Thus, one of the apparent discrepancies (see Sect. 1) between the \cite{Binney91} model and the CO \lvd ~ disappears. 
}

%%%%%%%%%%%%%%%%%%%%%%%%%%%%%%%%%%%%%%%%%%%%%%%%%
\section{Modeling the star counts map}
\subsection{The star counts map}
We have modeled the 2MASS star counts map of \cite{Alard01} to determine a realistic stellar potential.
In this section we summarize the methods and the main conclusions of Alard's work.

Using the H and K bands  and assuming that we know the intrinsic color of the sources ($H$ and $K$), one can correct for extinction one of the bands as follows:

\begin{equation}
m_K^{corr}=m_K- [ A_K/(A_H-A_k)][(m_H-m_K)-(H-K)] 
\end{equation}

where $m_{K,H}$ are the observed apparent magnitudes in K and H band (thus suffering extinction),  $A_{K,H}$ is the extinction in each band and  $m_K^{corr}$ is the extinction corrected magnitude in K band. At the distance of the bulge a cutoff $m_K=9$ (see below) implies that the sources are early M-giants (see Lopez-Corredoira et al. 2001). 
Assuming an intrinsic color  $H-K$ for M giants of  $\sim 0.17$ \citep{Binney98, Wainscoat92} and taking into account the relative extinctions $A_k/A_v=0.112$ and $A_h/A_v=0.175$ \citep{Rieke85}, one can write:

\begin{equation}
m_K^{corr}=m_K-1.77 \, (m_H-m_K)+0.30 
\end{equation}

\cite{Alard01} defined an extinction corrected magnitude $m_e\equiv m_K^{corr}-0.30$ as

\begin{equation}
m_e \equiv  m_K-1.77  \, (m_H-m_K)
\end{equation}
and constructed his star counts map using this magnitude with a cutoff of 9 mag. 
In most of this map, the density profile at constant longitude is exponential. This exponential profile is also present in numerical simulations. \cite{Combes90} showed that a disk with a small bulge in its center can evolve in a peanut shaped bar with a nearly exponential profile perpendicular to the plane of the galaxy.
However, \cite{Alard01} noticed that the observed density profile  in the inner four degrees of the Galaxy shows an excess of sources in the plane. Subtracting an exponential profile corresponding to disk and bulge, he found that the residual seems to be a small, lopsided,  nuclear bar in the galactic center. This structure has not been found in previous studies since the data lacked either the depth or the resolution.

\subsection{The model}

In this section we describe how we have modeled the 2MASS star counts map. 
The number of stars in a volume $dV$ located at $x$ with a magnitude in the interval $M$ and $M+dM$ is:

\begin{equation}
dN=\Phi(M,x)dMdV
\end{equation}

If the distribution of magnitudes is the same in all the space we can write $\Phi$ as
$\Phi(M,x)=\phi(M)\rho(x)$, where $\phi(M)$ is the {\it luminosity function} (the fraction of stars with magnitude $M$) and $\rho(x)$ is the { star density} in the point $x$.
Therefore,  the total number of stars with a magnitude $M$ lower than a cutoff  $M_c$ is:

\begin{equation}
N=\int_{M<M_c}\int_{v} dN=\int_{-\infty}^{M_c}\int_{\mathrm{V}} \phi(M) dM  \rho(x,y,z)dV
\end{equation}
 
As K band luminosity function $\phi(M)$, we have used the function given by  \cite{Wainscoat92}, which is very similar to that of \cite{Eaton84}. However,  we work with extinction-corrected apparent magnitudes $m$, which are related to the intrinsic magnitudes by:  

\begin{equation}
M= m+5-5 \log_{10} s
\end{equation}
where $s$ is the distance in pc. Therefore we actually deal with a function of the apparent magnitude and the distance $\phi'(m,s)$.

Our model of the density distribution is given by three components: a triaxial bulge or large bar, an exponential disk and a small nuclear bar. The exact functional form of each component is given below.

\subsubsection{Disc}
We use the exponential disc of \cite{Wainscoat92}, which is  defined as:

\begin{equation}
\label{eq:disc}
\rho^D(r,z)=
\rho_0 \exp\left(-\frac{(r-D)}{h_r}-\frac{|z|}{h_z}\right)  \\
\end{equation}

where $\rho_0$ stars number density in the solar neighborhood, $D$ is the distance from the sun to Galactic center, $h_z$ and $h_r$ are a vertical and radial scale parameters, respectively. 

%In some models we have also used and exponential disk with inner truncation (D=8.5 kpc) as defined by Lopez-Corredoira et al. ()
%\begin{equation}
%\left\lbrace
%\begin{array}{ll}
%\rho(r,z)=\rho_0 \exp\left(-\frac{(r-D)}{h_r}-\frac{|z|}{h_z}\right) & r>3.5 \mathrm{kpc} \\
%\rho(r,z)=\rho(3.5 kpc, z) \frac{r_{kpc}}{3.5} & r<3.5 \mathrm{kpc} \\
%\end{array}
%\right.  
%\end{equation}

\subsubsection{Triaxial bulge} 
From the COBE/DIRBE images we know the bulge of the Milky Way is {\it boxy} \citep{Dwek95}.
The 2MASS data confirm this result \citep{Lopez-Corredoira05}. Therefore, we have adopted a {\it boxy} Gaussian (Dwek et al. function G2) to represent the bulge. 

\begin{equation}
\label{eq:g2}
\rho^B(x',y',z')=\rho_0 \exp (-0.5r_s^2)
\end{equation}
\begin{equation}
r_s=\left( 
\left[ \left(\frac{x'}{x_0}\right)^2+\left(\frac{y'}{y_0}\right)^2\right]^2+
\left(\frac{z}{z'_0}\right)^4 \right)^{1/4}
\end{equation}

The bulge function is expressed in a coordinate system $(x',y',z')$ that follows its symmetry axes. Therefore, a rotation is needed to get the expression in the  $(x,y,z)$ coordinate system. 

\begin{equation}
\left\lbrace\begin{array}{l}
x'=x \cos\beta^B + y \sin \beta^B\\
y'=-x \sin \beta^B + y \cos \beta^B\\
z'=z\\
\end{array} \right.
\end{equation}

\begin{figure}
\begin{center}
\includegraphics[width=6cm]{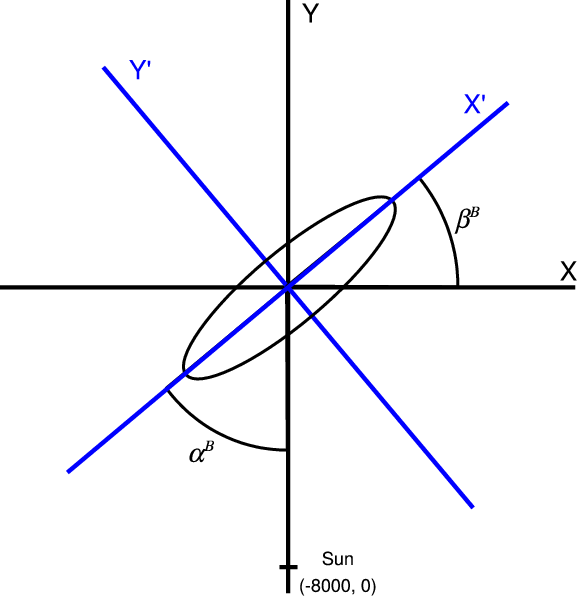}
\caption{ Sketch of the different coordinates systems and angles used to define the bars in the Galactic plane. The $z$ and $z'$ axis are coincident and the $z>0$ semiaxis is in the northern Galactic hemisphere.}
\label{fig:sisref}
\end{center}
\end{figure}

{Figure~\ref{fig:sisref} shows the $(x,y)$ and $(x',y')$ axis seen from the $z=z'>0$  hemisphere. The figure also shows the definition of the angle $\beta^B$, which is measured counterclockwise from the $x>0$ semi-axis. However, many papers measure
the inclination of the bar clockwise from the Sun-GC line to the near side of the bar.
 Therefore, for an easy comparison with previous results we have defined the angle $\alpha^B$ as shown in the figure.}

\subsubsection{Triaxial nuclear bar}
In contrast to the bulge, the small nuclear bar found by Alard does not seem to be boxy. Therefore we have taken a triaxial Gaussian function (Dwek et al. function G1) to represent this component:
 
\begin{equation}
\label{eq:g1}
\rho^b(x'',y'',z'')=\rho_0 \exp (-0.5r^2)
\end{equation}
\begin{equation}
r=\left[ \left(\frac{x''-x_1}{x_0}\right)^2+\left(\frac{y''}{y_0}\right)^2+\left(\frac{z''}{z_0}\right)^2 \right]^{1/2}
\end{equation}

Where $x_1$ allows to fit a lopsided bar.  To obtain the bar density in the $(x,y,z)$ coordinate system one should apply the rotation:

\begin{equation}
\left\lbrace\begin{array}{l}
x''=x \cos\beta^b + y \sin \beta^b\\
y''=-x \sin \beta^b + y \cos \beta^b\\
z''=z\\
\end{array} \right.
\end{equation}

{The definition of $\beta^b$ is exactly analogous to that of $\beta^B$ shown in Fig.~\ref{fig:sisref}. Again, in the following, and for coherence with most of the literature in the subject, we will use an angle $\alpha^b$ measured  clockwise from the Sun-GC line to refer to the inclination of the nuclear bar.}

\subsubsection{Final star counts equation}
Finally, we must also convert $\rho(x,y,z)$ to $\rho'(l,b,s)$ since we want to model the star counts in Galactic coordinates $(l,b)$ and to integrate along the line of sight ($s$).
 If $(x,y,z)$ is a right-handed reference frame centered in the galactic center, with the plane of the sky in the $xz$-plane with $x>0$ for $l<0$, and $z>0$ towards the Galactic north pole, and assuming that the sun is located in $(x,y,z)=(0,-D,0)$ then:

\begin{equation}
\label{eq:trans}
\begin{array}{l}
x=-s \cos(b) \sin(l)\\
y=-D+s\cos(b)cos(l)\\
z=s \sin(b)\\
\end{array}
\end{equation}

The star counts equation towards $(l_0,b_0)$ is then: 
\begin{equation}
\label{eq:starcounts}
\begin{array}{l}
N(l_0,b_0,m_c)  = \Delta l \, \Delta b \times  \\
 \times  \int_{m_{min}}^{m_{max}} dm \int_{0}^{s_{max}} ds\, \phi'(m,s)\, \rho'(l_0,b_0,s)\,  s^2 \cos b \\
\end{array}
\end{equation}

where $s^2 \cos b$ is the Jacobian of the coordinates transformation (Eq. \ref{eq:trans}). We have  assumed that the density is constant in the interval from $l_0$ to $l_0+\Delta l$ and from $b_0$ to $b_0+\Delta b$.

We have computed the integral in magnitudes from  $m_{min}=-10.0 \ \mathrm{mag}$  to $m_{max}=9.0 \ \mathrm{mag}$ with $\Delta m =0.5 \  \mathrm{mag}$.
The integral on the distance along the line of sight $s$ has been computed up to $s_{max}=20\ \mathrm{kpc}$ using a variable step $\Delta s$:

\begin{equation}
\Delta s = \left\lbrace  
\begin{array}{ll} 
0.002\ \mathrm{kpc} & s, R < 0.100\  \mathrm{kpc} \\
0.008\ \mathrm{kpc} & s, R < 0.300\ \mathrm{kpc} \\
0.010\ \mathrm{kpc} & s, R < 0.700\ \mathrm{kpc} \\
0.050\ \mathrm{kpc} & s, R < 1.500\ \mathrm{kpc} \\
0.100\ \mathrm{kpc} & s, R < 3.100\ \mathrm{kpc} \\
0.200\ \mathrm{kpc} & s, R < 6.300\ \mathrm{kpc} \\
0.300\ \mathrm{kpc} & s, R > 6.300\ \mathrm{kpc} \\
\end{array} \right.  
\end{equation}

%...................................................
\section{Fitting the star counts map}

We have fitted the star counts map using the model described above. 
This  model has 14 free parameters corresponding to the disc ($\rho_0, \, h_r,   \,h_z$), the bulge ($\rho_0, x_0, y_0, z_0, \alpha$) and the nuclear bar ($\rho_0, x_0, y_0, z_0, \alpha, x_1$).

The star counts equation is evaluated for an initial guess of the density function parameters.
A $\chi^2$ parameter is computed  as the difference of the star counts data points ($N_{data_i}$) and the model ($N_{model_i}$).
\begin{equation}
\chi^2= \frac{1}{n}  \sum_{i=1}^n \frac{(N_{data_i}-N_{model_i})^2}{\sigma^2_i} 
\end{equation}

where  $n$ is the number of points to be fitted and $\sigma$ is the error of the data points. 
We have estimated the $\sigma$ from Fig. 1 of \cite{Alard01} as the dispersion of the star counts with respect to an exponential vertical profile. Afterwards we have used  an iterative process to minimize $\chi^2$ using the Levenberg-Marquardt algorithm \citep{Press92}. 

\subsection{Fitting the disc and the boxy bulge}

\begin{figure}
\vspace{0cm}
\centerline{\includegraphics*[ angle=-90,width=9cm]{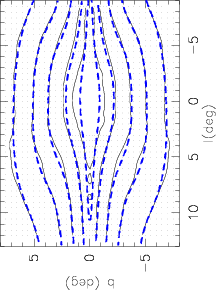}}
\caption{Thin solid contours represent the star counts map. Thick dashed lines are the best fit obtained with a star counts model with $\alpha^B=20\deg$  and $h_r=2.5$ kpc. The points show the data set used to fit disc and bulge.}
\label{fig:res-Bd}
\end{figure}

 We have not attempted to fit all the free parameters at once. 
First, we have fitted a star counts model with the disc and the boxy bulge to a reduced data set
consisting of one point every 10 pixels (one data point every $0.5\deg$). 
In the disc region, we have taken one point every 5 pixels ($0.25\deg$) across the disc.
Finally, we have neglected the data points in the inner region ($-3 < l (deg) < 3, \ -2 <b (deg) <2$) since they are clearly dominated by the nuclear bar.
This reduced data set contains 1494 points and it is shown in Fig. \ref{fig:res-Bd}.
The errors of these data points used to calculate $\chi^2$ has been estimated to be 15 $\%$  as explained above.

%\begin{figure}
%\centerline{\includegraphics*[bb=304 50 533 363, angle=-90,width=9cm]{f_map_m9_bulge.png}}
%\caption{Data set used to do a first fit of the disc and the boxy bulge}
%\label{fig:data1}
%\end{figure}

We have fitted the data for fixed values of the bulge angle ($\alpha^B=60, 45, 35, 20, 10, 0$ deg) and the disc radial scale ($h_r=1.5, 2, 2.5, 3.5$ kpc). The results obtained for the other free parameters are shown in Table \ref{tab:res-Bd}. 
For all $\alpha^B$'s, the best fits are always obtained with $h_r$=2.5 kpc.
This is in agreement with the radial scale length of the disc derived from previous studies in the infrared, which have found values in the range from 1.9 \citep[2MASS][]{Lopez-Corredoira05} to 2.6 \citep[COBE][]{Freudenreich98}.
Regarding the angle of the bulge, the best fit is obtained for $10\deg$, although $\chi^2$ is only a 2.8 $\%$ higher for $\alpha^B=20\deg$ (Figure \ref{fig:res-Bd} shows the best fit with  $\alpha^B=20\deg$). This is also in agreement with other studies of the structure of the inner Galaxy.
\cite{Dwek95} derived angles in the range $50\deg-74\deg$ depending on the function assumed to represent the bulge (they found $74\deg$ with the same boxy Gaussian that we use here), while \cite{Freudenreich98} derived angles in the range of $75\deg-81\deg$ also using COBE/DIRBE data. 
A high angle of $78\deg$ has also been measured from the 2 Microns Galaxy Survey (TMGS) by \cite{Lopez-Corredoira00}.
On the other hand, the work of \cite{Lopez-Corredoira05} using 2MASS data found $\alpha^B=20\deg-35\deg$.

The shape of the bulge as given by the best fit with $\alpha^B=10\deg$ is $x_0/x_0:y_0/x_0:z_0/x_0=1:0.5:0.3$ while the best fit with $\alpha^B=20\deg$ has axis ratios of $1:0.55:0.4$ which are in perfect agreement with those derived by
\cite{Lopez-Corredoira05} also with 2MASS data or by \cite{Lopez-Corredoira00} with TMGS data. In contrast, other studies have favored somewhat thinner bulges with ratios of $1:0.3-0.4:0.3$ \citep{Dwek95, Freudenreich98, Bissantz03}.

%\vspace{1cm}
%With  $\alpha=70$  $h_r=1.0$...
%{\includegraphics[angle=-90,width=9cm]{f_hr6.png}}

\begin{table*}
\begin{center}
\caption{Results of the fits to the star counts data of Fig. \ref{fig:res-Bd} using an exponential disc and a triaxial boxy-Gaussian bulge}
\label{tab:res-Bd}
\begin{tabular}{lllll|lll|l}
\hline
\multicolumn{5}{c|}{Bulge} & \multicolumn{3}{c|}{Disk} & \\
 $x_0$&$y_0$&$z_0$&$\rho_0$&$\alpha^B$&$h_r$&$h_z$& $\rho_0$ & $\chi^2$ \\
   kpc & kpc & kpc && deg   &   kpc & kpc  & &  \\
\hline
 1.068& 0.498& 0.346&0.389E-02&  0.0& 2.500& 0.168&0.249E-03&0.412\\
\hline                                                                                                                               
 1.045& 0.508& 0.348&0.452E-02&  10.0& 3.000& 0.142&0.365E-03&0.290\\
 1.117& 0.513& 0.362&0.383E-02&  10.0& 2.500& 0.153&0.254E-03&0.259\\
 1.243& 0.540& 0.383&0.303E-02&  10.0& 2.000& 0.161&0.145E-03&0.264\\
\hline                                                                                                                            
 0.817& 0.491& 0.349&0.615E-02&  20.0& 3.500& 0.134&0.469E-03&0.336\\
 0.840& 0.486& 0.359&0.557E-02&  20.0& 3.000& 0.144&0.366E-03&0.288\\
 0.884& 0.489& 0.375&0.477E-02&  20.0& 2.500& 0.156&0.255E-03&0.266\\
 0.975& 0.513& 0.400&0.376E-02&  20.0& 2.000& 0.166&0.145E-03&0.284\\
 1.118& 0.574& 0.432&0.283E-02&  20.0& 1.500& 0.169&0.551E-04&0.325\\
\hline                                                                                                                               
 0.703& 0.444& 0.355&0.768E-02&  35.0& 3.500& 0.135&0.470E-03&0.335\\
 0.716& 0.435& 0.366&0.708E-02&  35.0& 3.000& 0.145&0.366E-03&0.291\\
 0.744& 0.429& 0.382&0.621E-02&  35.0& 2.500& 0.158&0.255E-03&0.274\\
 0.812& 0.442& 0.408&0.499E-02&  35.0& 2.000& 0.169&0.145E-03&0.301\\
 0.937& 0.496& 0.445&0.371E-02&  35.0& 1.500& 0.172&0.551E-04&0.350\\
\hline                                                                                                                           
 0.670& 0.381& 0.369&0.851E-02&  45.0& 3.000& 0.146&0.367E-03&0.293\\
 0.693& 0.367& 0.385&0.767E-02&  45.0& 2.500& 0.159&0.255E-03&0.278\\
 0.752& 0.365& 0.412&0.637E-02&  45.0& 2.000& 0.171&0.145E-03&0.309\\
\hline                                                                                                                             
 0.621& 0.200& 0.373&0.171E-01&  60.0& 3.000& 0.147&0.367E-03&0.297\\
 0.637& 0.132& 0.389&0.226E-01&  60.0& 2.500& 0.161&0.254E-03&0.286\\
 0.681& 0.179& 0.416&0.140E-01&  60.0& 2.000& 0.172&0.144E-03&0.327\\
\hline
\end{tabular}
\end{center}
\end{table*}

\subsection{Fitting  disc, bulge and nuclear bar}
\begin{figure*}
\vspace{0cm}
\centerline{\includegraphics*[ angle=-90,width=15cm]{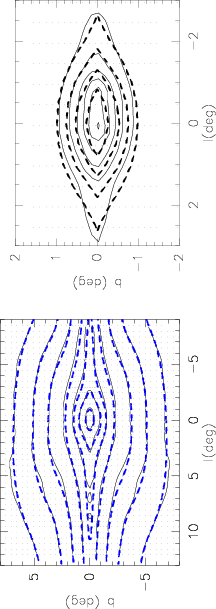}}
\centerline{\includegraphics*[angle=-90,width=15cm]{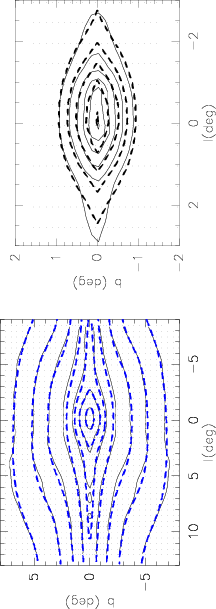}}
\caption{{\it Upper panels:} Thin solid contours represent the star counts map. Thick dashed lines are the best fit obtained with a star counts model with $\alpha^B=10\deg$, $h_r=2.5$ kpc and $\alpha^b=75$ deg. The panel on the right is a zoom of the nuclear region. The points indicate the data used to do the fit.
{\it Lower panels:} Same that the upper panels for the best fit with $\alpha^B=45\deg$,  and a lopsided bar with $\alpha^b=60\deg$.}
\label{fig:res-Bdb}
\end{figure*}

In the second data set we have also  included the star counts in the innermost region of the Galaxy 
($-3 < l < 3, \ -2 <b <2$). 
We have taken one point every 3 pixels ($0.15\deg$) across the disc. 
The data set is shown in Fig. \ref{fig:res-Bdb}. It contains 1734 points.
From Fig. 1 of \cite{Alard01} is clear that the dispersion of the star counts map in the nuclear region is smaller than in the bulge.  We have estimated a $\sigma$ of $5 \%$ for these data points.

%\begin{figure}
%\centerline{\includegraphics*[bb=304 50 533 363, angle=-90,width=9cm]{f_map_m9_all.png}}
%\caption{Data set used to do fit the disc, the bulge and the nuclear bar}
%\label{fig:data2}
%\end{figure}

We have fitted the data using our star counts model with three components: disc, bulge and nuclear bar.
We have taken as starting point the best fits for $\alpha^B$=10$\deg$,  20$\deg$,  35$\deg$,  45$\deg$  and 60$\deg$ derived in the previous section (all the parameters of the disc and the bulge are fixed except the factors $\rho_0$).
The best fit results for the parameters of the nuclear bar are summarized in Table \ref{tab:res-Bdb}.
The full set of results for the nuclear bar fits are listed in 
Tables \ref{tab:res-Bdb1} - \ref{tab:res-Bdb4}.

The best fits  are found with a thick triaxial bar with typical axis ratios of $1:(0.7-0.8):(0.5-0.6)$ and $\alpha^b$ in the interval $60-120$ {\it  for all the $\alpha^B$'s}.
This implies that the parameters of the nuclear bar can indeed be constrained independently of the bulge and that $\alpha^b \neq \alpha^B$.
Globally, the best fit is obtained with $\alpha^B=10\deg$  and a nuclear bar with  $\alpha^b=90\deg-105\deg$. some examples of good fits are shown in Fig. \ref{fig:res-Bdb}.

\begin{table*}[]
\caption{Summary of the parameters of the nuclear bar as derived from the best fits to the 2MASS star counts data. The first row are fits with $x_1$ fixed to 0 and the second row corresponds to fits with free $x_1$ to allow for lopsidedness. The full set of results as a function of the orientation of the bulge $\alpha^B$ are listed in Tables \ref{tab:res-Bdb1} to \ref{tab:res-Bdb4}.
}
\label{tab:res-Bdb}
\begin{center}
\begin{tabular}{llllll|l}
\hline
 $x_0^b$ & $y_0^b$ & $z_0^b$ &$\rho_0^b$ & $\alpha^b $ & $x_1$ &   $\chi^2$ \\
  kpc     & kpc      &   kpc   &              & deg        & kpc      &           \\
\hline
 0.141-0.150& 0.130-0.138& 0.095-0.097 & 0.141-0.150 & 60-120 &  0.0        & 0.357-0.376\\
 0.158-0.166& 0.115-0.133& 0.088-0.094 & 0.151-0.178 & 60-120 & 0.022-0.025& 0.331-0.346 \\
\hline                  
\end{tabular}           
\end{center}            
\end{table*}

%\subsection{Fitting  disc, bulge and {\it lopsided} nuclear bar}

 Finally, we have also performed fits with a lopsided nuclear bar ($x_1 \neq 0$).
The results are shown in Tables \ref{tab:res-Bdb1} - \ref{tab:res-Bdb4} and they are 
summarized in Table \ref{tab:res-Bdb}.
The best fits are still obtained with a nuclear bar that is almost perpendicular to the Sun-GC line.  
The symmetry center of the bar would be shifted by 22-25 pc towards the third Galactic quadrant if $\alpha^b=60\deg-90\deg$ while it would be shifted by the same quantity but towards the fourth quadrant if  $\alpha^b=90\deg-120\deg$.

%%%%%%%%%%%%%%%%%%%%%%%%%%%%%%%%%%%%%%%%%%%%%%%%%%%%%%
\section{Simulations of the gas dynamics: methods}

We have used the code of \cite{Combes85} to simulate the motion of gas clouds in the potential obtained from the mass distribution inferred with the star counts model.
This code has already been used successfully to understand the gas distribution and dynamics of a number of galaxies as 
%M51 \citep{Garcia-Burillo93}, 
M100  \citep{Garcia-Burillo98}, 
%NGC 4321 \citep{Sempere95} 
or NGC 4736 \citep{Gerin91, Mulder96}.

\subsection {Stellar potential, rotation curve and masses}

The potential is obtained from the mass distribution derived in the previous section, using standard 
FFT techniques in a Cartesian grid of  $512 \times 512 \times 16$ cells.
{ The cells are cubic with a linear size of 58.6 pc, therefore the physical size of the grid 
is $30 \times 30 \times 0.94$ kpc$^3$. This is good compromise to model the large scale 
dynamics at least to a radius of 10 kpc and to have a good enough spatial resolution to study the Galactic center.

As described by \cite{Combes81}, the interaction between periodically reproduced 
images is avoided following the scheme of \cite{Hohl69}, i.e. the potential computations
are done in a grid 8 times bigger, of $1024 \times 1024 \times 32$ cells.
 The disk mass is truncated at 15kpc in radius, and the mass distribution 
is assumed spherical outside, so that its influence in the center is cancelled.
This is equivalent to assume that a spherical dark halo is dominating the mass outside.
This assumption should have a negligible influence on the dynamics of the inner Galaxy, 
which is the main interest of this paper.
}

The stellar potential is  decomposed  in its axisymmetric and non-axisymmetric parts. The axisymmetric potential is calculated in radial bins using the averaged mass in the bin. The non-axisymmetric part of the potential is calculated with a mass distribution corresponding to the difference between the total-mass and the axisymmetric distribution.

The stellar number density derived in the previous section should be multiplied by a factor $f$ to obtain the total mass density. 
This factor accounts for the mass of the stars detected by 2MASS but also for the mass of those stars that are not counted in the 2MASS maps.
{ In addition, we have to introduce a dark halo to explain the observed velocities at large radii.} The total density used is then:

\begin{equation}
\label{eq:rhot}
\rho^T=f \ ( \rho^D + \rho^B +  \rho^b ) + \rho^H
\end{equation}

where the halo density is defined by a Plummer law:

\begin{equation}
\label{eq:halo}
\rho^H(r)=\frac{\rho_0^H}{[1+(r/r_c)^2]^{5/2}}
\end{equation}

\begin{figure}
\centerline{\includegraphics[angle=-90,width=8.5cm]{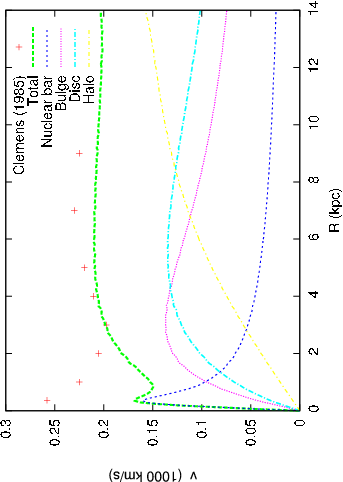}}
\caption{Rotation curve obtained with the halo and the scaling discussed in Sect. 5.1 for a bulge and nuclear bar orientation of $\alpha^B=20\deg$ and $\alpha^b=75\deg$, respectively. The different lines show the contribution of the different mass components to the total rotation curve. The red points indicate the rotation curve measured from observational data.
The mass of the halo , disc, and bulge are $M^H=7\,10^{10}$~M$_\odot$, $M^D=2.9\,10^{10}$~M$_\odot$, $M^B=1.9\,10^{10}$~M$_\odot$, respectively. The mass of the nuclear bar is $M^b=2.0\,10^{9}$~M$_\odot$.}
\label{fig:rotcur1}
\end{figure}

{
To determine the scaling factors $f$ and $\rho_0^H$ we have compared the terminal velocities and the rotation curve of the models with those derived from spectroscopic observations  \citep{Clemens85, Fich89, Burton93}.  
To obtain terminal velocities of the right magnitude (see Figs. \ref{fig:simu_a70} and \ref{fig:simu_o30}) and a flat rotation curve at large radii (Fig. \ref{fig:rotcur1})},
one should apply a scaling factor $f=3.9 \ 10^8$ and a halo with  $ \rho_0^H= 1.12\,10^5$ and    $r_c=15.0 \, \mathrm{kpc}$. 
With these factors and the  parameters given in Tables \ref{tab:res-Bdb1} - \ref{tab:res-Bdb4}, the function $\rho^T$ gives the total mass density in units of $10^3$~\Msun/kpc$^3$.
Integrating in the total grid size, the typical masses of the different  components are: $M^H=7\,10^{10}$~M$_\odot$, $M^D=2.9\,10^{10}$~M$_\odot$, $M^B=1.9\,10^{10}$~M$_\odot$ and $M^b=2.0\,10^{9}$~M$_\odot$, which are in good agreement with previous determinations \citep[see for instance][]{Weiner99, Fux99}.

Figure \ref{fig:rotcur1} shows  an example of rotation curve. 
The modeled curve compares well with the rotation curve derived from observations of the interstellar gas for radius larger than 2 kpc.
The agreement seems worse for small radii.
However the comparison in this region is not straightforward. 
On the one hand, the measured curve has been derived using the tangent point method assuming that the clouds move in circular orbits, which is not true in the innermost region.
In addition, this curve depends on the position of the observer in the Galaxy.
On the other hand, the modeled rotation curve has been derived with the axisymmetrical potential and with an azimuthal average of the non-axisymmetrical potential.
This rotation curve is a kind of average curve independent of the position of the observer.

%\begin{figure}
%\centerline{\includegraphics[angle=-90,width=7cm]{f_rotcur_0.png}}
%\caption{Rotation curve. $\rho^T=f ( \rho^D +  \rho^B + \rho^b )$ }
%\label{fig:rotcur}
%\end{figure}

\begin{figure}
\centerline{\includegraphics*[ angle=-90, width=8cm]{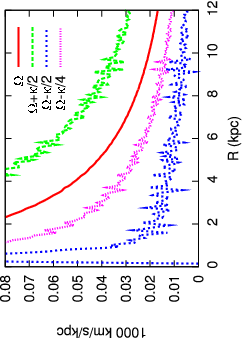}}
\caption{Angular frequencies as a function of radius in the epicyclic approximation for the potential discussed in Sect. 5.1}                       
\label{fig:omega}
\end{figure}

Figure \ref{fig:omega} gives the rotation velocity $\Omega$ as a function of galactocentric radius.
It also shows the $\Omega + \kappa/m$ curves (where $\kappa$ is the epicyclic frequency).
For a given bar pattern speed $\Omega_p$, these curves give the location of the corotation ($\Omega=\Omega_p$), the Outer 
Lindblad resonance ($\Omega+\kappa/2=\Omega_p$), the Inner Lindblad resonances ($\Omega-\kappa/2=\Omega_p$) 
and the Ultra Harmonic Resonance ($\Omega-\kappa/4=\Omega_p$).

\subsection{Gas cloud dynamics}

The gas clouds are modeled as particles that are initially launched in  the axisymmetric potential with a velocity corresponding to the circular velocity of the potential at the radius of the particle.
The initial radial distribution is an exponential disk with radial scale of 2 kpc. 
%Its distribution perpendicular to the plane is Gaussian.
The non-axisymmetric part of the potential is then introduced gradually multiplying by  the factor $1-\exp[-t/\tau]^2$ with a time delay of $\tau=$100 Myr. The angular speed of the non-axisymmetrical pattern, $\Omega_p$, is constant along the simulation.

We do not use a mass spectrum for the clouds and we do not simulate star formation. 
%All the clouds have a constant mass of 3.1\,$10^4$ M$_\odot$. 
The simulations contain typically $10^6$ particles. 
The effects of the gas self-gravitation are neglected.
The clouds move as test particles in the stellar potential except when they collide inelastically.
The forces due to the stellar potential are evaluated every 1 Myr and the particles are moved according to them.
The particles are followed in three dimensions but the collisions are computed in a  two-dimensional grid with 480 $\times$ 480 cells.
Clouds in the same grid cell can collide. The relative velocity of two collisions partners loses 60$\%$ of its absolute value in the collision. A  detailed discussion of the collisional schema can be found in \cite{Combes85} and \cite{Casoli82}.

%%%%%%%%%%%%%%%%%%%%%%%%%%%%%%%%%%%%%%%%%%%%%%%%%%%%%%%%%%%%%%%%%%
\section{Simulations results and discussion: large scale structure}

 Figures \ref{fig:simu_a70} and \ref{fig:simu_o30} show face-on views and \lvd \ for a grid of models with $\alpha^B$ from $10 \deg$ to $45 \deg$   and pattern speeds from 20 to 50 \kmskpc.
For an easy comparison with the observations, we have overlaid in the \lvd s  the TVC points, the boxes indicating the tanget points and the solid and dashed lines shown in Fig. \ref{fig:dame}.
We have also plotted the four logarithmic spirals of \cite{Nakanishi06} in the face-on views. 
The models shown in Figs. \ref{fig:simu_a70}  and \ref{fig:simu_o30} exhibit a number of common  characteristics.
The galaxies have two major spiral arms but they also exhibit a second pair of less prominent spiral arms.
In the inner $\sim 4$ kpc, there are arms that contour the bar forming a kind of oval (hereafter lateral arms).
The density of gas inside this oval decreases down to a radius of $\sim 1$ kpc, where there is a ring.
This ring is connected to the lateral arms by the off-axis arms (or dustlanes).
In the inner hundreds of parsecs there is a nuclear condensation of gas.

{ The Milky Way galaxy is a very complex system and up to now there is not a single simulation of the gas dynamics that can explain all the observed features, nor quantitatively nor qualitatively. 
\cite{Englmaier99} have compared their models to the terminal velocity curve, they have also compared the position the arms tangent points in the CO \lvd , at $|l|=30\deg$  and $50 \deg$. 
The two \lvd s  presented by \cite{Fux99} has been selected by the global resemblance to the \lvd s obtained from CO and HI data. They reproduce qualitatively the 3-kpc arm and the Connecting Arm.
\cite{Weiner99} constrained  their models by comparing with the HI terminal velocity curve.  
Finally, \cite{Bissantz03} have used mainly the terminal velocity curve and some voids in the observed \lvd \ to compare with their simulations. 
To select the best models in the present work we have compared with:  \textit{i)} the terminal velocity curve, \textit{ii)} the inclination of the GMR in the \lvd , \textit{iii)} the existence or not of the 3kpc arm and when it exists, its inclination and absolute velocity 
  \textit{iv)} the arms tangent points \textit{v)} the shape and size of the HI 1kpc ring.
}

\subsection{Pattern speed and orientation of the bulge}

\begin{figure*}[h]
\centerline{\includegraphics[angle=-90, width=14cm]{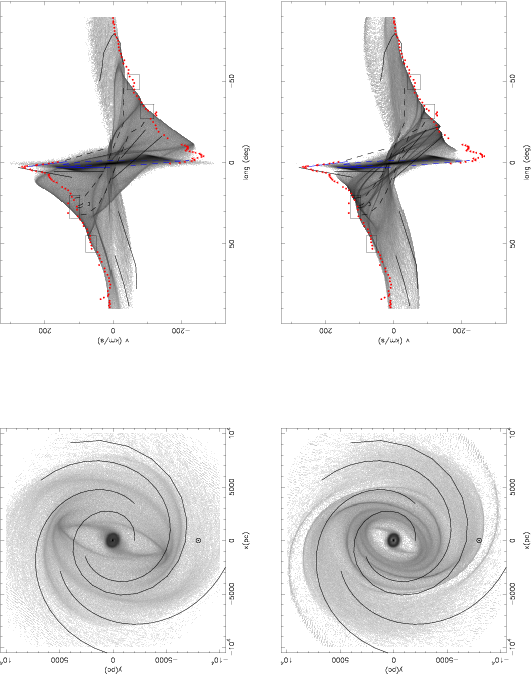}}

\centerline{\includegraphics[  angle=-90, width=14cm]{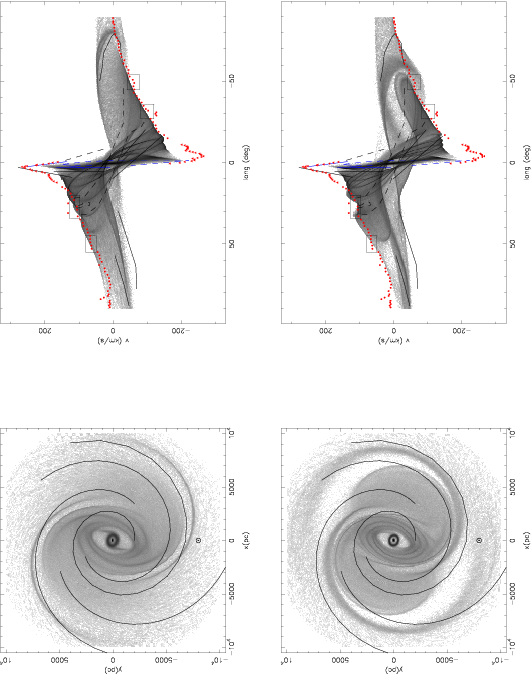}}

\caption{Simulations results for $\alpha^B=20\deg$. The different rows represent the face-on view and the longitude velocity diagram for different pattern speeds of 20, 30, 40 and 50 km/s/kpc (from upper to lower rows). The solid lines represent the spiral arms as defined by \cite{Nakanishi06}. The line Sun-GC first crosses the Sgr-Carina arm and the Scutum-Crux arm. On the other direction, the line of sight to the Galactic anticenter crosses the Perseus arm. The fourth arm is the Norma-Outer arm.}                       
\label{fig:simu_a70}
\end{figure*}

{ Figure \ref{fig:simu_a70} shows face-on views and lv-diagrams for a grid of models with $\alpha^B=20\deg$  and pattern speeds from 20 to 50 \kmskpc.
The model with $\Omega_p=20$ \kmskpc \ reproduces some spiral tangent points and has an inner lateral arm with the inclination of the 3-kpc arm, however it does not reproduce the terminal velocity curve nor the velocity of the 3-kpc arm, for instance.
On the other extremity, the model with $\Omega_p=50$ \kmskpc \ does not reproduce satisfactorily neither the terminal velocity curve nor the GMR nor the 3-kpc arm.

In contrast, the  models with $\Omega_p=30$ and $40$ \kmskpc \ give a very good overall representation of the Galaxy.
Both reproduce satisfactorily the spiral tangent points at $|l|=30, 50 \deg$   within  $\pm 5 \deg$ and $\pm 10$ \kms, approximately.
In addition, the model with $\Omega_p=40$ \kmskpc \ reproduces very well the Carina arm and its tangent point at $\approx -80 \deg $. 
However, the locus of the inner spiral arms and the 3-kpc arm in the modeled \lvd~ are too steep in comparison with the observed \lvd. 
In contrast, the  GMR and the the 3-kpc arm is very well reproduced from $|l|=90\deg$ to $15\deg$  by the model with $\Omega_p=30$ \kmskpc. 
In particular the 3-kpc arm is reproduced with both the good inclination and the velocity within $\sim 10$ \kms.
In addition, the  terminal velocity curve for $l>5 \deg$ is very well reproduced.
Given the difficulty of the task, it will not be realistic to give a very narrow interval for the bar pattern speed, from our best models \textit{we conclude that the most probable value for $\Omega_p$ should be in the range $30-40$ \kmskpc}.
However, since this work is  devoted to the inner Galaxy structure and kinematics in the following we will mainly discuss models with $\Omega_p = 30$ \kmskpc \ since they explain the  terminal velocity curve, the 3-kpc arm and the GMR better than the model with $\Omega_p  = 40$ \kmskpc.
} 

\begin{figure*}[h]
\centerline{\includegraphics[  angle=-90, width=14cm]{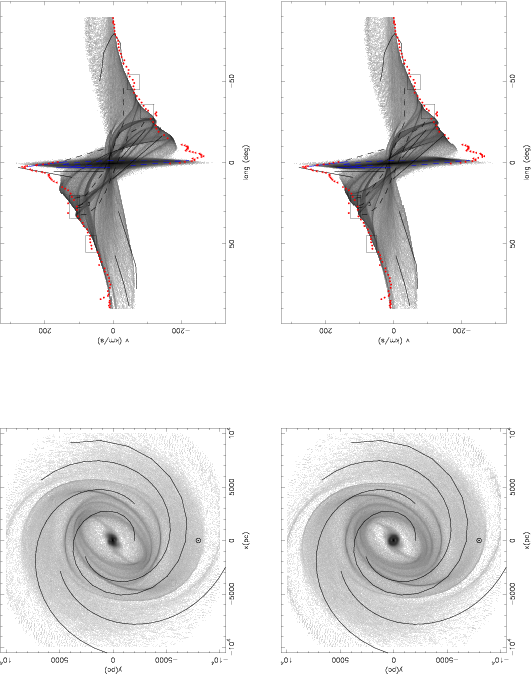}}

\centerline{\includegraphics[  angle=-90, width=14cm]{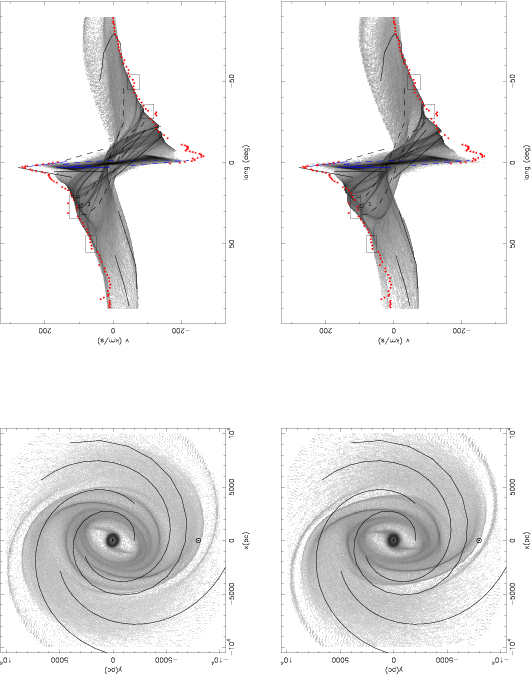}}

\caption{Simulations results for $\Omega_p=30\deg$. The different rows represent the face-on view and the longitude velocity diagram for different orientations of the bulge $\alpha^B$  of 45, 35, 20 and 10 deg from the upper to the lower row}                       
\label{fig:simu_o30}
\end{figure*}

Figure \ref{fig:simu_o30} shows face-on views and \lvd s for a grid of models with constant $\Omega=30$ \kmskpc \ and $\alpha^B$'s of 45$\deg$ , 35$\deg$ , 20$\deg$  and 10$\deg$ ($\alpha^b$ is constant to 90$\deg$  for all the models).
For $\alpha^B=10\deg$, the velocity along the line of sight is too low to explain the high non-circular velocities of the 3 kpc arm ($-53$ \kms \ at $l=0\deg$) and the observed arm is not reproduced in the $lv$ diagram.
In contrast, the 3-kpc arm and the inclination of the GMR is well reproduced for $\alpha^B$ from 20$\deg$ to 45$\deg$.
{
Nevertheless, the model with $\alpha^B=45\deg$ does not reproduce correctly the positive velocity terminal curve. In addition, there is no ring at 1 kpc in the face on view.

Once again, we do not pretend to give a very narrow interval of angles for the inclination angle of the bar since the analysis, as in other published works, is still rather qualitative.
We reckon that \textit{the inclination of the bar should be in the range from 20$\deg$ to 35$\deg$}, with some preference for 20$\deg$ since the quantitative agreement of the 3-kpc arm and the terminal velocity curve is better. 
Therefore, the model with $\Omega_p = 30$ \kmskpc \ and $\alpha^B=20\deg$ will be considered as our \textit{standard} model in the rest of the paper.

\subsubsection{Comparison with previous works}
The inclination of the bar derived from our models is in good agreement with previous determinations \citep[][see also Sect. 1]{Binney97, Fux99, Lopez-Corredoira05}. 
Regarding the bar pattern speed, our simulations give results that are lower than those of 
\cite{Englmaier99} or  \cite{Bissantz03}, which suggest $\Omega_p=50-60$ \kmskpc, but in good agreement with the results of \cite{Weiner99}, $\Omega_p=42$ \kmskpc, or \cite{Fux99}.
In his self-consistent gas and stars simulations, Fux found that the pattern speed evolves from 50 to 30 \kmskpc .
Taking into account our potential, a pattern speed of 30 or 40 \kmskpc \ gives a corotation radius of 7 and 5 kpc, respectively (see Fig. \ref{fig:omega}).
Therefore, the corotation radius is well beyond the bar extremity, which is the characteristic of a \textit{slow} bar \citep[see for instance][]{Sellwood88}.

To explain the fact that the dustlanes in the inner spiral arms of some galaxies are located in the concave section of the arms, and assuming that the bars are fast (i.e., they extend to the corotation), \cite{Sellwood88} proposed the existence of different speeds for the bar and the spiral patterns. 
Indeed, for the Milky Way some works have derived low rotations speeds of 13-20 \kmskpc \ for the spiral pattern \citep{Lin69, Morgan90, Amaral97}.
Gas flow models with independent speeds for the bar and the spiral patterns have been presented by \cite{Bissantz03}, who discussed that different speeds for the spiral and the bar patterns could explain some regions devoid of gas in the \lvd .
However, the overall agreement of their  \lvd s with the observations is not better than other models like that of \cite{Fux99} or our models, which are able to reproduce \textit{quantitatively} features as the 3-kpc arm.

%Although a discussion of an independent pattern speed for the spirals is beyond the scope of this paper, 
It is interesting to remark that our simulations favor a relatively slow bar and not a fast bar as assumed by \cite{Sellwood88}.
Furthermore, the pattern speed of the relatively slow bars models \citep[30-40 \kmskpc,][this work]{Weiner99, Fux99}   is comparable to 
several determinations of the spiral pattern speed that give values of 30-35 \kmskpc \ \citep{Mishurov99,Fernandez01, Lepine01}. 
%The determination of the spiral pattern speed by \cite{Debattista02} using observations of OH/IR stars gives even higher values of 59 \kmskpc. 
On the other hand, \cite{Ibata95} have measured a rotation speed of $25\pm 4$~\kmskpc \ for the bulge in the $0.7<R<3.5$ kpc region.
 Therefore, at present there are still many uncertainties on the speeds of the different patterns and it is not clear whether the spirals and the bar are decoupled or they turn at the same speeds. }
{  Simulations with an independent pattern speed for the spirals would be needed to study  this question in detail but this is beyond the scope of the current paper, which is mainly devoted to the inner Galaxy.}

%%%%%%%%%%%%%%%%%%%%%%%%%%%%%%%%%%%%%%%%%%%%%%%%%%%%%%
\subsection{The GMR}

\begin{figure*}[h]
\centerline{\includegraphics*[ angle=-90, width=16cm]{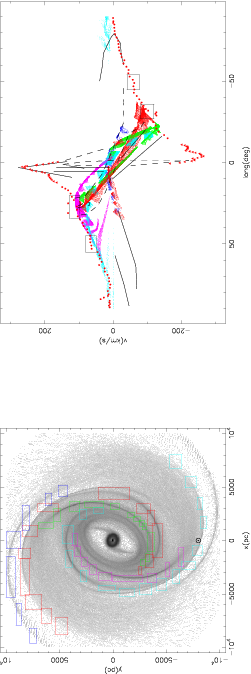}}

\centerline{\includegraphics*[angle=-90, width=16cm]{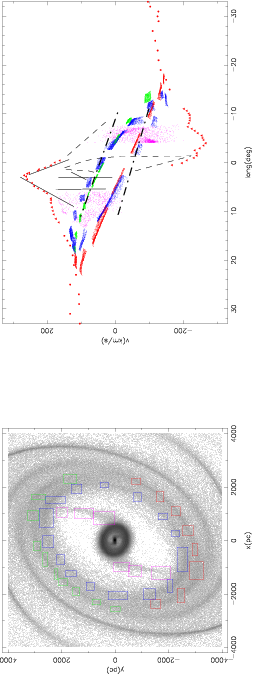}}

\caption{Simulations results for $\alpha^B$=20$\deg$ and $\Omega_p=30$. We have selected different regions in the face on views to identify the locus of the different structures in the \lvd \   (using the same color code in the face on view and the \lvd). Solid and dashed lines, circles and triangles are the same that in the previous figures. The dot-dashed lines in the lower \lvd \ are the fits to the near and far 3-kpc arms as given by \cite{Dame08b}}                       
\label{fig:simu_a70_sche}
\end{figure*}

To get further insight in the structure and the dynamics of the Galaxy, we present below a more detailed analysis of the model with
$\Omega_p=30$ \kmskpc \ and $\alpha^ B$=20$\deg$.
We have selected different structures in the face-on view and we show the locus of the different structures in the \lvd ~ (Fig. \ref{fig:simu_a70_sche}).

{
As already mentioned in Sect. 2.2, it is uncertain whether the  GMR is an actual ring or it is composed by imbricated spiral arms.
In our models the radius of the GMR (4-5 kpc) corresponds exactly with the position of the Ultra-Harmonic resonance for $\Omega \sim 30$ \kmskpc  \ (Fig. \ref{fig:omega}).  
However, we do not support the idea that the GMR is a resonant ring. 
}

As shown in the upper panels of Fig. \ref{fig:simu_a70_sche}, in our models the GMR is composed by the inner parts of the spiral arms, which at those radius show almost circular motion.
This is in agreement with the face-on views of the CO and HI data of \cite{Nakanishi06}, who have proposed that the GMR is formed by the Scutum-Crux arm (near side of the GMR) and the inner part of the Sgr-Carina (far side of the GMR).
{\cite{Englmaier99} discussed that if formed by imbricated arms, the GMR should be outside the corotation assuming that there cannot exist significant circular orbits inside the corotation to produce a feature resembling a circular ring.
However, this statement only holds for fast bars, for which all the region inside the corotation is dominated by the bar. 
In contrast, in our models with a relatively slow bar, the inner part of the spiral arms is inside the corotation radius and exhibit almost circular motions. 
}

\subsection{The 3-kpc  arm and its far side counterpart}

{
Lower panels of Fig. \ref{fig:simu_a70_sche} show the locus of the lateral arms that contour the bar.
The two dot-dashed lines in the \lvd \ are the fit to the 3-kpc arm and to its far-side counterpart as given by \cite{Dame08b}.
As already mentioned, the locus of the lateral arms in the \lvd \ reproduce quite well the structure of the 3-kpc arm.
This is in agreement with \cite{Fux99}, who has already proposed that the 3-kpc arm could be a lateral arm.
On the other hand, \cite{Fux99}  proposed that the ``135 \kms \ arm'' is the far side counterpart of the 3-kpc arm.
However, our models predict a far side counterpart of the 3-kpc arm whose locus in the \lvd \ is almost symmetrical to the 3-kpc arm.
This is in perfect agreement with the recently found far 3-kpc arm \citep{Dame08b}.
The full implications of the new finding by \cite{Dame08b} will be presented elsewhere.

The fact that the 3-kpc arms are lateral arms imply that they cannot be a channel to transport gas from the corotation to the CMZ as suggested by  \cite{Habing06}, at least not directly.
}

\subsection{The off-axis arms or dustlanes}
In contrast to the 3-kpc arms, the dustlanes do contribute to the transport of gas to the inner regions. In our models the locus of the dustlanes in the \lvd \ are vertical structures with velocities from $\sim 0$ \kms \ to more than 100 \kms. The maximum velocity increases as $l$ decreases.
These vertical structures in the \lvd\ resemble the Clumps found by \cite{Bania77} and discussed recently by \cite{Liszt06, Liszt08}.
Nevertheless, our models do not reproduce the Galactic longitude of the observed Clumps (5.5$\deg$ and 3.1$\deg$), 
we confirm that Banias's Clumps are probably shocked gas in the dustlanes, which was the original interpretation of \cite{Stark86} for the ``Clump 2".
The fact that the Clumps are composed of gas suffering the dustlanes shocks is also suggested by observational evidences both kinematical \citep[linewidths, velocity gradients,][]{Stark86, Liszt06} and chemical \citep{Rodriguez-Fernandez06}.
Taking into account the discrete appearance of the Bania's clumps at two Galactic longitudes, it is likely that the dust lanes of the Milky Way are patchy as it is commonly observed in external galaxies. 

\cite{Fux99} and \cite{Marshall08} have proposed a slightly different interpretation.
They propose that the Connecting Arm is the locus of the near side dustlane in the \lvd. 
However, the linewidths of the gas in the Connecting Arm do not show the velocity dispersion expected in the dustlanes shocks \citep[see for instance][]{Reynaud1998}.
Indeed probably both the Bania's clumps and the Connecting Arm are related to the dustlane.
We reckon that Bania's Clumps are clouds that are actually suffering the strong shocks expected in the dustlanes while the Connecting Arm is most likely the post-shocked and accelerated gas \citep[see also the discussion in][]{Liszt06}.  
Indeed, in our simulations  the maximum velocity of the gas in the dustlanes increases as $l$ decreases with the same steep of the Connecting Arm.
Observationally, this interpretation is supported by the fact that the Clump at $l=5.5\deg$ seems to be connected to to Connecting Arm at a negative latitude \citep[see][]{Bitran97}. The situation was less clear for the ``Clump 2" at $l=3.1\deg$ since it is found at positive latitudes.
{ However, this apparent problem has been solved with the recent discovery by \cite{Liszt08} of an emission feature with the same inclination in the \lvd \ that the Connecting Arm but at positive latitude.
In addition it seems connected to the Clump 2. }
In any case, the Clump 2 exhibits a rich shock chemistry as expected for the gas in the dustlanes shocks \citep{Rodriguez-Fernandez06}

%%%%%%%%%%%%%%%%%%%%%%%%%%%%%%%%%%%%%%%%%%%%%%%%%%%%%%
\section{Simulations results and discussion: nuclear bar, HI ring and CMZ}

Figure \ref{fig:simu_b} shows the face-on view of the inner 2 kpc of the modeled galaxies and the \lvd s for different orientations of the nuclear bar. All the other parameters are the same that those discussed in the previous section and the large scale face on views and \lvd s
are the same independently of the orientation of the nuclear bar.

The dustlanes end in a ring and inside this ring there is a small bar-like structure that is connected to the ring by two small spiral arms.
The HI ring has inner and outer radius of 300 pc and 800 pc, respectively, with are in excellent agreement with the HI observations of \cite{Liszt80} (see their figure 3).
The ring is almost circular. In some simulations we have also found elliptical rings with the major axis approximately perpendicular to the large bar. Therefore, the ring is not supported by x1 orbits elongated in the direction of the large bar  \citep[see a review of the different proposed models in][]{Ferriere07}.

\subsection{The orientation of the nuclear bar}

In the central hundreds of parsecs there is an elongated structure that corresponds to the gas response to the nuclear bar and that resembles the observed CMZ.

\begin{figure*}
\centerline{\includegraphics[  angle=-90, width=14cm]{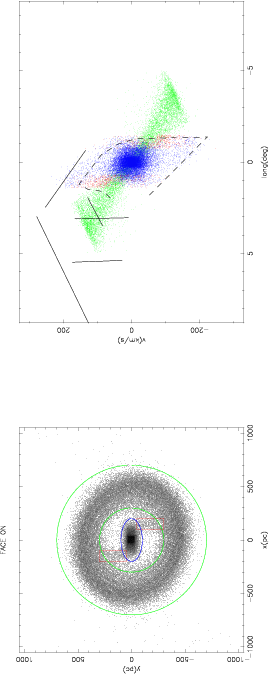}}

\centerline{\includegraphics[ angle=-90, width=14cm]{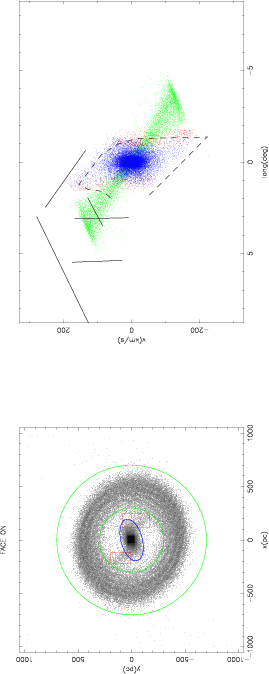}}

\centerline{\includegraphics[  angle=-90, width=14cm]{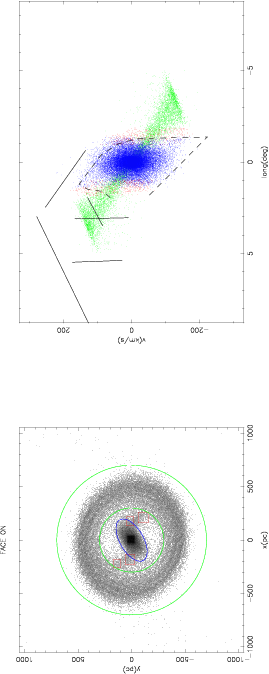}}

\centerline{\includegraphics[ angle=-90, width=14cm]{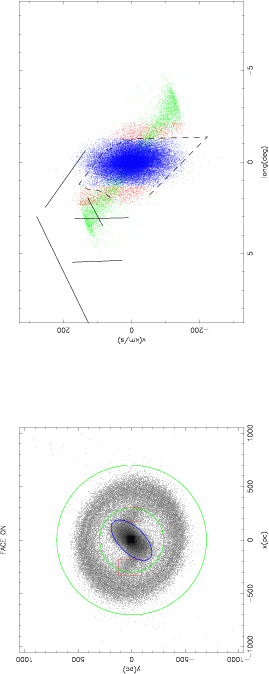}}

\caption{Simulations results for $\alpha^B=20\deg$, $\Omega_p=30$ and $\alpha^b$=90$\deg$, 75$\deg$, 60$\deg$, and $45\deg$ (from top to bottom).}                       
\label{fig:simu_b}
\end{figure*}

We have analyzed the shape of the \lvd\ as a function of the nuclear bar orientation.
Figure \ref{fig:simu_b} shows the \lvd\ obtained for  nuclear bar orientations from 0$\deg$, to 15$\deg$, 30$\deg$ and 45$\deg$.
Our modeled \lvd s do not reproduce the components K and J of \cite{Rodriguez-Fernandez06}.
In contrast, the \lvd\ of the CMZ is a kind of parallelogram that resembles the observed one (Fig. \ref{fig:bally}).
The inclination of the observed parallelogram is very well reproduced for an angle of $15\deg$.
Little after the \cite{Binney91} paper, \cite{Jenkins94} tried to reproduce the parallelogram discussed by \cite{Binney91} using n-body simulations but with little success. 
To our knowledge, this is the first time that numerical simulations of the gas dynamics of the Milky Way reproduce naturally the parallelogram of the \lvd\ of the CMZ.
This is probably due to the accurate potential that we have determined with the deep 2MASS data.
 
It is worth noting that the orientation angle of the small bar inferred from our simulations is in excellent agreement with our models of the 2MASS star counts and with the face on view of the CMZ inferred from CO and OH data by \cite{Sawada04}.
We conclude that the observed CMZ is most likely the gas response to the nuclear bar and that the orientation of the nuclear bar is $\alpha^b \sim 75\deg$.

\subsection{The mass of the nuclear bar}
\begin{figure}
\centerline{\includegraphics[angle=-90,width=8.5cm]{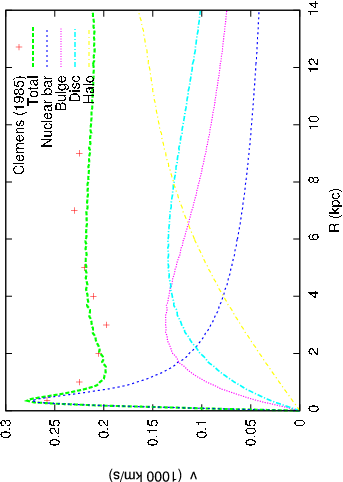}}
\caption{Same that Fig. \ref{fig:rotcur1} but with a nuclear bar  mass  of $M^b=5.5\,10^{9}$~M$_\odot$.}
\label{fig:rotcur2}
\end{figure}

\begin{figure*}
\centerline{\includegraphics*[angle=-90, width=14cm]{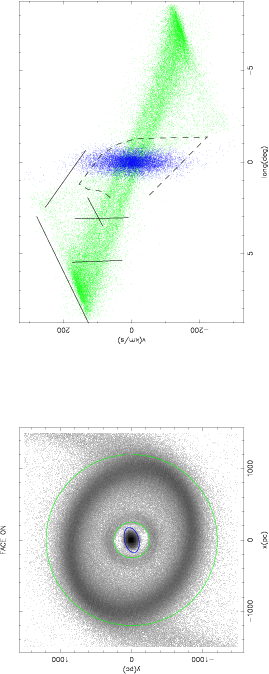}}
\caption{Simulations results for $\alpha^B=20\deg$, $\Omega_p=30$ and $\alpha^b=75\deg$  (with nuclear bar mass of 25$\%$ of bulge mass)}                       
\label{fig:simu_mass}
\end{figure*}

In the simulations discussed above, the velocity dispersion of the CMZ is well reproduced for a mass of the nuclear bar of $2.0\,10^{9}$ M$_\odot$ (Sect. 5.1).
However, the terminal velocities of the HI ring do not match the observed ones. 
To get further insight on the nuclear bar mass we have increased its value in order to reproduce the terminal velocities of the HI ring.
 In these models, 
 the shape of the different mass components is still given by the fits to the 2MASS star counts map  (Tables \ref{tab:res-Bdb1} - \ref{tab:res-Bdb4}). However, we have considered different scaling factors for the nuclear bar and for the bulge and disc. Thus, instead of Eq. \ref{eq:rhot}, the total density is given by:

\begin{equation}
\label{eq:rhot2}
\rho^T=f \ ( \rho^D + \rho^B) + f^b \ \rho^b  + \rho^H
\end{equation}

where $\rho^H$ is given by Eq. \ref{eq:halo}.
Therefore, with respect to previous models, the only parameter that we change  is $f^b$.
Figure \ref{fig:rotcur2} shows the rotation curve for $f^b=1.1 \, 10^9$, which gives a
nuclear bar mass of $5.5 10^{9}$ M$_\odot$. All the other parameters and masses are the same that those mentioned in Sect.~5.1.
The simulations results for the inner 2 kpc of the Galaxy are shown in Fig. \ref{fig:simu_mass}. 
The face on view is characterized by a large ring with inner and outer radius of 200 and 1200 pc, respectively.
The \lvd \ of this ring reproduces the terminal velocity of the HI observations, which
 at positive longitude coincides  with the Connecting Arm.
Indeed, \cite{Liszt80} proposed that the Connecting Arm could be the edge of the HI ring.

It is difficult to reproduce both the terminal velocities of CMZ and HI ring, however from the previous considerations we conclude that the mass of the nuclear bar should be in the range $(2-5.5)\,10^{9}$ ~M$_\odot$.

\subsection{Is the nuclear bar lopsided? }
The nuclear bar in the star counts map of \cite{Alard01} seems to be shifted towards negative longitudes (in contrast to the gas distribution that seems to be shifted towards positive longitudes). 
Our star counts model shows that fits with a lopsided nuclear bar have a $\chi^2$ that is 9$\%$ lower than the $\chi^2$ for models that do not allow a lopsided bar. The symmetry center of the bar could be shifted by 22-25 pc with respect to the dynamical center of the Galaxy.

We have studied the gas response to a lopsided nuclear bar in order to investigate whether the observed asymmetry of the CMZ could be explained by a lopsided nuclear bar as inferred from the 2MASS data. 
As discussed by \cite{Morris96} the signature of a $m=1$ mode in the Galactic center would be a shift of the gas distribution in $l$ but also a shift of the velocity centroid of the \lvd.

\begin{figure*}
\centerline{\includegraphics[ angle=-90, width=14cm]{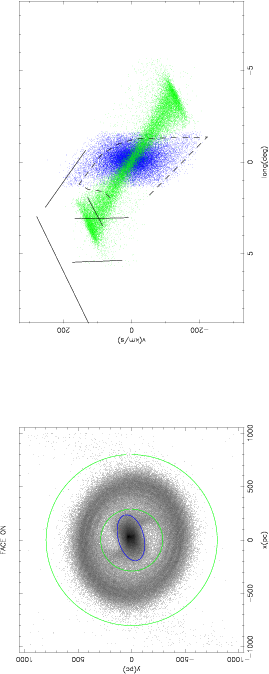}}

\centerline{\includegraphics[ angle=-90, width=14cm]{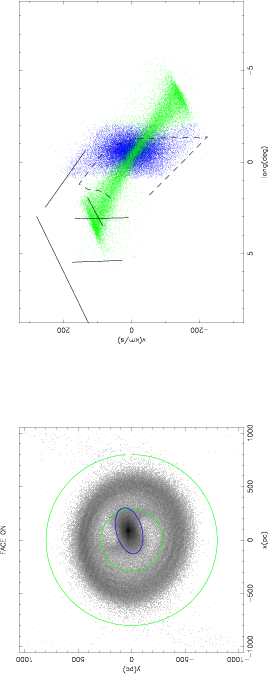}}

\centerline{\includegraphics[ angle=-90, width=14cm]{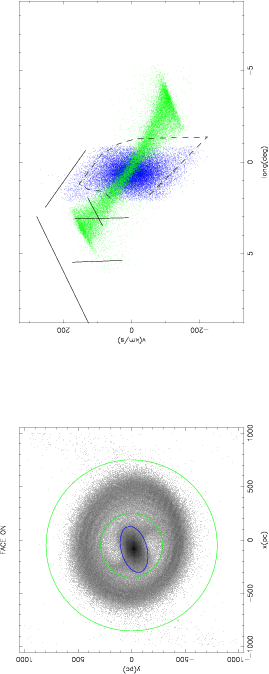}}
\caption{Simulations results for $\alpha^B=20\deg$, $\Omega_p=30$ and $\alpha^b=75\deg$. The upper panel corresponds to the best fit to the 2MASS data with a lopsided bar ($x_1=23$ pc). The middle panel is a simulation with the same nuclear bar but $x_1=90$ pc and the lower panel is a simulation with the same nuclear bar but $x_1=-9$ }                       
\label{fig:simu_lop}
\end{figure*}

Figure \ref{fig:simu_lop} shows a simulation of the gas flow in the potential computed from the fit to the 2MASS star counts map with $\alpha^b=75\deg$ (for which $x_1=0.023$ kpc).
The \lvd ~ of the gas exhibit a parallelogram whose sides show the good inclination.
The simulated parallelogram is only slightly asymmetric in $l$ and $v$.
To better understand the effect of a lopsided nuclear bar, we have also computed a simulation with the same nuclear bar but artificially shifted by 90 pc along its major axis towards the third Galactic quadrant ($x_1=0.090$ kpc). The results are also shown in Fig. \ref{fig:simu_lop}.
The \lvd\ is still only slightly asymmetric in $v$ but now it is clearly asymmetric in $l$. The gas response to the lopsided stellar bar  follows the star distribution and it is lopsided towards negative longitudes as well.
To obtain a \lvd ~ with a parallelogram similar to that of the CO data shown in Fig. \ref{fig:bally}, the nuclear bar should be lopsided towards the first quadrant.
This is shown in  Fig. \ref{fig:simu_lop}, where $x_1=-0.090$ kpc.
However, such a bar is not compatible with the 2MASS star counts. 

We conclude that, both from the fits to the 2MASS data and from the numerical simulations of the gas dynamics, there are no clear evidences of an intrinsic lopsidedness in the stellar potential. The observed asymmetry of the gas distribution in the CMZ cannot be explained as the gas response to a lopsided nuclear bar in the way suggested by Alard's map.

%%%%%%%%%%%%%%%%%%%%%%%%%%%%%%%%%%%%%%%%%%%%%%%%%
\section{Discussion}
\subsection{The origin of the asymmetry of the CMZ}

Although the origin of the  asymmetry of the CMZ is a long-standing problem, it has not been discussed much in the literature. Two exceptions are the papers by \cite{Combes96} and \cite{Morris96}. The first paper proposed that the observed asymmetry could well be due to a $m=1$ mode of the potential since  lopsidedness is commonly observed in the central
regions of external galaxies. The second paper discusses how the effect of the presence of a $m=1$ mode in the Galactic center would be an asymmetry of the CMZ both in longitude and in velocity.
However, we have shown in the previous section that the possible lopsidedness of the stellar distribution cannot explain the observed asymmetry of the CMZ.
Furthermore, there is not a clear evidence of an intrinsic asymmetry of the stellar potential.
Regarding the gas, the  distribution is clearly asymmetric in $l$ but
the parallelogram of Fig. \ref{fig:bally} is rather symmetric in $v$.
In addition,  the structures found by \cite{Sofue95} that seem to be the signature of a quasi-circular ring inside the CMZ and that could be the nuclear bar equivalent to the HI ring, do not show neither an asymmetry in $v$. 

Indeed our simulations propose an alternative explanation for the asymmetric distribution of the CMZ in which the stellar potential in the GC is strictly symmetric. 
The simulations show that the HI ring can be connected to the CMZ by a pair of small spiral arms.
For instance, the simulation with $\alpha^b=75$ (second row of Fig. \ref{fig:simu_b}) shows that the clouds in the arm seen at negative longitudes will be seen in the \lvd \ outside of the observed parallelogram while the clouds in the arm seen at
positive longitudes  ``fill" the observed parallelogram at the position of the {\it l=1.3$\deg$-complex}.
Observations of the molecular gas in external galaxies show that, in contrast with most numerical simulations, the gas distribution in real galaxies is rarely symmetric.
This could also be the case in the Milky Way. 
The apparent asymmetry of the CMZ can be due to material falling into the CMZ from the HI ring through only one of the inner spiral arms.
In this context, strong shocks are expected in the interaction region, that would be the 
{\it l=1.3$\deg$-complex}.

Observationally, the {\it l=1.3$\deg$-complex} looks very different to the Sgr's cloud complexes.
For instance, in the $lb$ map of Fig. \ref{fig:bally}, it is clear that the gas distribution  from Sgr C to Sgr B is  rather symmetric and thin.
The asymmetry in $l$ is indeed exclusively due to the {\it l=1.3$\deg$-complex}.
Furthermore, the {\it l=1.3$\deg$-complex} shows a very special latitude extension, much higher that the rest of the CMZ.
The velocity structure of this cloud complex is also quite singular, showing a large velocity dispersion (the {\it l=1.3$\deg$-complex} is indeed the most prominent feature in the \lvd). 
The chemistry gives us also interesting hints on the nature of this cloud complex.
On the one hand, \cite{Huttemeister98} have found that the highest  abundances  of the shock tracer molecule SiO,  are found at one of the extremities of the CMZ: in the {\it l=1.3$\deg$-complex}.
They have already proposed that this cloud complex is the scenario of strong shocks due to gas falling into the CMZ.
On the other hand, \cite{Rodriguez-Fernandez06} have presented a more complete study of the SiO emission in all the kinematical structures of the GC.
They have detected SiO  in at least one cloud of every kinematical structure.
However, the only feature where they detected high abundances of SiO in {\it every} observed cloud is the {\it l=1.3$\deg$-complex}. In agreement with \cite{Huttemeister98}, \cite{Rodriguez-Fernandez06} have measured the highest SiO abundances in this region.
Therefore, the $lb$ distribution, the velocity structure, and the chemistry, imply that the cloud complex giving rise to the observed asymmetry of the CMZ (the {\it l=1.3$\deg$-complex}) shows significant differences to other CMZ complexes. 
In particular, it shows the signature of strong shocks.
Therefore, we reckon that the observed asymmetry of the CMZ can well be the result of gas falling into the CMZ and not due to an intrinsic asymmetry of the stellar potential.
This explanation is also supported by our detailed simulations of the stellar structure and the gas dynamics.

\subsection{The nature of the nuclear bar}
{
Double bars are commonly observed in external galaxies \citep{Shaw93,  Wozniak95, Elmegreen96}. 
Around 28 $\%$ of barred galaxies have a secondary bar \citep{Erwin02, Laine02}. 
Nuclear or \textit{secondary} bars are objects of prime interest to explain the gas inflow towards the center of a galaxy and the fueling of Active Galactic Nuclei \citep{Shlosman89}.

Several works have studied the stellar orbits that support nested bars \citep{Maciejewski97, Maciejewski00} and their stability   \citep{El-Zant03}.
In addition, double barred galaxies have been the subject of a number of numerical simulations of the star and gas dynamics
that have shown a plethora of phenomena as: 
\textit{i)} secondary bars forming in purely stellar discs
\citep{Rautiainen02} or only in the presence of gas \citep{Combes94}.
\textit{ii)}   secondary bars forming before  \citep{Rautiainen99} or after the primary bars \citep{Combes94, Englmaier04, Heller07a},  
\textit{iii)} decoupled secondary bars rotating slower \citep{Heller01} or faster \citep{Friedli93, Combes94, Rautiainen02, Englmaier04} than the primary bar,
\textit{iv)} two misaligned bars rotating at the same speed  \citep{Shaw93, Combes94}. %, Englmaier04}
 
% \citep{Friedli93, Shaw93, Combes94, Heller01, Maciejewski02, Rautiainen02, Shlosman02, Englmaier04, Debattista07}.
%Recently, \cite{Heller07a} have done the first simulations of the formation of nested bars from cosmological initial conditions. 

%All these simulations have shown a plethora of possible scenarios as: 

%One of the most interesting aspects of nested bars is their coupling.  Some secondary bars are found to rotate with the same pattern speed than the primary bar \citep{Shaw93, Combes94} while others  rotate at with a much larger speed \citep{Friedli93, Combes94, Rautiainen02}.  

The secondary bars rotating at the same speed than the primary bars have been explained by \cite{Shaw93} and \cite{Combes94} as follows: the gas clouds tend to shift from the closed x1 orbits to the perpendicular x2 orbits of the primary bar. However,   dissipative collisions between the gas clouds reduce the orthogonality of this phase shift and the gas settles in a leading phase-shifted bar. If the gas fraction is high, its gravitational influence is then sufficient to modify the stellar component and to form a secondary bar, that is still coupled to the primary bar. 
{If  secondary and  primary bar remain coupled and they continue rotating with the same speed, the bars will align or form an angle of 90$\deg$ since the primary bar exerts a gravitational torque on the secondary bar and vice-versa \citep{Friedli93}.
Alternatively, the secondary bar may increase its rotation speed.  
Thus, the coupled phase can be a precursor of a future decoupled phase. The dynamical process from coupled to decoupled gaseous nuclear bars have been studied by \cite{Heller01}, \cite{Englmaier04}
and, more recently, by \cite{Heller07a}, who have done the first simulations of the formation of nested bars from cosmological initial conditions. } In these simulations a secondary bar forms in response to the gas inflow along the stellar primary bar. This gaseous bar is initially corotating with the primary bar. Subsequent mass inflow strengthens this bar giving rise to even more rapid gas inflow. Along this process the secondary bar shrinks size and the pattern speed increases inversely proportional to the bar size. 
The pattern speeds of the two bars are such that the secondary bar corotation coincides with the primary bar Inner Lindblad Resonance, suggesting that non-linear interactions between the two bars are at play \citep{Tagger87}. 

Our simulations are the very first attempt to model the Milky Way with two nested bars and have been done assuming a common pattern speed for both bars. Therefore, we cannot study in detail the  coupling and the evolution of the nuclear bar.
Nevertheless, it is worth noting that these simulations explain many characteristics of Galaxy at scales from the disk to the nuclear region, some for the first time as the parallelogram of the CMZ.
Therefore, the observational data are \textit{compatible} with a scenario of coupled bars rotating with the same speed.
Our results imply  that the nuclear bar is leading the large primary bar by $\sim 55\deg$, as expected in this context \citep{Combes94}.
Thus, the decoupling of the nuclear bar can still not be effective in the Milky Way.
The dynamical decoupling of the two bars  will be the subject of a forthcoming paper.
}

%%%%%%%%%%%%%
\section{Conclusions}
We have presented gas flow models in the mass distribution derived from 2MASS
star counts using a model with three components (disk, bulge and nuclear bar).
Our dynamical models are the first ones that include a central mass cusp (the nuclear bar) constrained by observations.
For the first time, we have obtained good models of the Milky Way from the scales of the spiral arms to the Central Molecular Zone (CMZ) in the Galactic center (GC).
The best models are found for  a bulge orientation  of $20-35\deg$ with respect to the Sun-GC line and a pattern speed of 30-40 \kmskpc. 
Which places the corotation radius at 5-7 kpc.
The simulations reproduce:
\begin{itemize} 
\item the spiral arms, giving in particular the good tangent point for the Carina arm.
\item the Galactic Molecular Ring (GMR), which is not an actual ring but the inner parts of the spiral arms and it is found approximately at the  position of the ultra-harmonic resonance
\item { the 3-kpc arm and its far-side counterpart, which are the lateral arms that surrounds the bar}
\item the HI ring, with a very similar size to that derived from the observations
\item the Central Molecular Zone (CMZ)
\end{itemize}
In addition,  we postulate the existence of small spiral arms arising from the extremities of the CMZ and linking the HI ring.

In our simulations, the CMZ is the gas response to the nuclear bar.
These simulations reproduce, for the first time,  the parallelogram shape of the \lvd ~ of the CMZ.
Using this shape we have been able to constrain the nuclear bar orientation, which is of $\sim 60\deg-75\deg$ with respect to the Sun-GC line, in excellent agreement with the 2MASS fits and the results of \cite{Sawada04}.
We have also studied the observed asymmetry of the CMZ and we have concluded that it  cannot be due to the lopsidedness of the nuclear bar suggested by the 2MASS maps. Gas dynamical simulations in a lopsided potential cannot
reproduce the observations, and in any case, we do not see a strong direct evidence in 2MASS data
of an intrinsic lopsidedness of the stellar potential.

We propose an alternative scenario to explain the observed asymmetry of the CMZ.
The asymmetry can be due to gas falling into the CMZ trough the inner spiral arm  in the {\it l=1.3$\deg$-complex} region.
We have also discussed that the {\it l=1.3$\deg$-complex} shows  significant differences with the, otherwise symmetric, CMZ that extends from Sgr C to Sgr B.
In particular, the  {\it l=1.3$\deg$-complex} shows all the signatures of the shocks expected if it is an interaction region where new material is falling into the CMZ.

{
In our models both bars rotate with the same pattern speed. The success of these models in explaining many characteristics of the Milky Way as the Central Molecular Zone (CMZ) imply that the observations are compatible with that assumption. Furthermore, in our models the  nuclear bar is leading the large bar by $\sim 55 \deg$ as expected for coupled nested bars \citep{Combes94}.
Therefore, in spite of the high mass of the nuclear bar  (10-25 $\%$ of the mass of the bulge) the dynamical decoupling my still not be effective. 
However, this question requires further studies and gas flow models with two pattern speeds or self-consistent gas and stars dynamics simulations.}

\begin{acknowledgements}
{ We thank Katia Ferri\`ere for a critical reading of the manuscript and Tom Dame for sharing with us their finding of the far-side counterpart of the 3-kpc arm before publication. 
We also thank an anonymous referee for useful comments that have improved the paper, in particular regarding the phenomena of several pattern speeds and nested bars in other galaxies.
}
The early stages of this work have benefited from support by a Marie Curie Fellowship of the European Community program ''Improving Human Research Potential and the Socio-economic
Knowledge base'' under contract number HPMF-CT-2002-01677.
\end{acknowledgements}

%   FLOATS  ---------------------------------------------------------------------

\clearpage

\begin{table*}[h!]
\caption{Results of the fits to the star counts data of Fig. \ref{fig:res-Bdb} using disc, bulge and  nuclear bar. The following parameters have been taken from Table \ref{tab:res-Bd} and kept fix: $\alpha^{B}=10, \ x_0^B=1.117,\ y_0^B=0.513, \ z_0^B=0.362  \ h_r=2.5 , \ h_z=0.153$. The bar angle $\alpha^b$ is also fixed, the other parameters are free and used to obtain the best fit to the data.
The first set of results have been obtained with $x_1$ fixed to 0. In the second set of results $x_1$ is a free parameter to fit.}
\label{tab:res-Bdb1}
\begin{center}
\begin{tabular}{ll|llllll|l}
\hline
Bulge      & Disk   &\multicolumn{6}{c|}{Bar}                     &                   \\
$\rho_0^B$  &$\rho_0^D$  &  $x_0^b$ & $y_0^b$ & $z_0^b$ &$\rho_0^b$ & $\alpha^b $ &  $x_1$&  $\chi^2$ \\
                &                & kpc     & kpc      &   kpc   &              & deg      &kpc        &           \\
\hline
%0.394E-02&0.269E-03& 0.141& 0.011& 0.024&0.279E+01&   0.0& 1.609\\
%0.395E-02&0.267E-03& 0.147& 0.006& 0.027&0.498E+01&  15.0& 1.442\\
%0.396E-02&0.261E-03& 0.168& 0.015& 0.038&0.169E+01&  30.0& 0.911\\
%0.392E-02&0.256E-03& 0.206& 0.049& 0.059&0.364E+00&  45.0& 0.465\\
%0.379E-02&0.256E-03& 0.224& 0.131& 0.103&0.106E+00&  60.0& 0.371\\
%0.372E-02&0.258E-03& 0.231& 0.152& 0.123&0.816E-01&  75.0& 0.438\\
%0.368E-02&0.258E-03& 1.133& 0.167& 0.125&0.155E-01&  90.0& 0.450\\
%0.378E-02&0.258E-03& 0.128& 0.145& 0.109&0.152E+00& 105.0& 0.396\\
%0.383E-02&0.255E-03& 0.215& 0.138& 0.092&0.110E+00& 120.0& 0.350\\
%0.392E-02&0.256E-03& 0.208& 0.053& 0.060&0.335E+00& 135.0& 0.449\\
%0.395E-02&0.260E-03& 0.170& 0.018& 0.040&0.136E+01& 150.0& 0.848\\
%0.394E-02&0.257E-03& 0.156& 0.037& 0.051&0.679E+00& 165.0& 0.582\\
%\hline                                                             
0.379E-02&0.257E-03& 0.195& 0.146& 0.103&0.103& 165.0&  0.0 & 0.374\\
0.380E-02&0.256E-03& 0.189& 0.144& 0.102&0.113& 150.0&  0.0 & 0.367\\
0.381E-02&0.256E-03& 0.175& 0.138& 0.100&0.129& 135.0&  0.0 & 0.362\\
0.381E-02&0.256E-03& 0.164& 0.131& 0.098&0.146& 120.0&  0.0 & 0.358\\
0.381E-02&0.255E-03& 0.161& 0.130& 0.097&0.151& 105.0&  0.0 & 0.357\\
0.381E-02&0.255E-03& 0.161& 0.130& 0.097&0.153&  90.0& 0.0  & 0.357\\
0.381E-02&0.256E-03& 0.161& 0.130& 0.097&0.151&  75.0&  0.0 & 0.358\\
0.381E-02&0.256E-03& 0.163& 0.131& 0.098&0.146&  60.0&   0.0& 0.359\\
0.380E-02&0.256E-03& 0.175& 0.137& 0.100&0.129&  45.0&   0.0& 0.363\\
0.380E-02&0.256E-03& 0.189& 0.144& 0.102&0.113&  30.0&  0.0 & 0.369\\
0.380E-02&0.257E-03& 0.192& 0.145& 0.103&0.106&  15.0&  0.0 & 0.375\\
0.380E-02&0.257E-03& 0.184& 0.141& 0.101&0.110&   0.0&  0.0 & 0.379\\
\hline
0.381E-02&0.256E-03& 0.267& 0.141& 0.099&0.081& 165.0& 0.080&0.331\\ 
0.381E-02&0.256E-03& 0.209& 0.130& 0.096&0.115& 150.0& 0.042&0.327\\
0.382E-02&0.255E-03& 0.180& 0.129& 0.092&0.141& 135.0& 0.030&0.324\\
0.384E-02&0.255E-03& 0.163& 0.128& 0.090&0.155& 120.0& 0.025&0.321\\
0.383E-02&0.255E-03& 0.155& 0.127& 0.093&0.160& 105.0& 0.023&0.322\\
0.382E-02&0.256E-03& 0.154& 0.116& 0.094&0.177&  90.0& 0.022&0.323\\
0.383E-02&0.255E-03& 0.158& 0.115& 0.093&0.175&  75.0& 0.023&0.322\\
0.383E-02&0.255E-03& 0.162& 0.129& 0.092&0.153&  60.0& 0.025&0.321\\
0.383E-02&0.255E-03& 0.177& 0.128& 0.091&0.142&  45.0& 0.031&0.321\\
0.383E-02&0.255E-03& 0.208& 0.133& 0.093&0.116&  30.0& 0.045&0.322\\
0.382E-02&0.256E-03& 0.274& 0.146& 0.097&0.080&  15.0& 0.090&0.326\\
\hline                  
\end{tabular}           
\end{center}

\end{table*}     
\begin{table*}          
\begin{center}          
\caption{Same that Table  \ref{tab:res-Bdb1} but with $\alpha^B=20, \ x_0^B=0.884,\ y_0^B=0.489, \ z_0^B=0.375  \ h_r=2.5 , \ h_z=0.156$. The first set of results have been obtained with $x_1$ fixed to 0. In the second set of results $x_1$ is a free parameter to fit.}
\label{tab:res-Bdb2}
\begin{tabular}{ll|llllll|l}                                                     
\hline
Bulge   & Disk   &\multicolumn{6}{c|}{Bar}              &                        \\
$\rho_0^B$  &$\rho_0^D$  &  $x_0^b$ & $y_0^b$ & $z_0^b$ & $\rho_0^b$ & $\alpha^b $  &  $x_1$ & $\chi^2$ \\
        &        & kpc    & kpc   &   kpc &         & deg     & kpc   &           \\
%\hline
%0.492E-02&0.267E-03& 0.146&-0.007& 0.028&0.464E+01&   0.0& 1.442\\
%0.491E-02&0.257E-03& 0.163& 0.040& 0.052&0.634E+00&  15.0& 0.600\\
%0.486E-02&0.255E-03& 0.180& 0.021& 0.071&0.958E+00&  30.0& 0.407\\
%0.488E-02&0.256E-03& 0.214& 0.056& 0.063&0.315E+00&  45.0& 0.459\\
%0.476E-02&0.256E-03& 0.251& 0.117& 0.097&0.110E+00&  60.0& 0.368\\
%0.465E-02&0.257E-03& 0.240& 0.155& 0.123&0.794E-01&  75.0& 0.440\\
%0.459E-02&0.257E-03& 1.132& 0.168& 0.125&0.159E-01&  90.0& 0.447\\
%0.466E-02&0.257E-03& 0.277& 0.156& 0.120&0.710E-01& 105.0& 0.425\\
%0.476E-02&0.255E-03& 0.257& 0.116& 0.097&0.110E+00& 120.0& 0.362\\
%0.488E-02&0.256E-03&-0.208& 0.058& 0.065&0.306E+00& 135.0& 0.440\\
%0.492E-02&0.258E-03&-0.171& 0.024& 0.049&0.966E+00& 150.0& 0.646\\
%0.492E-02&0.259E-03& 0.160& 0.006& 0.046&0.432E+01& 165.0& 0.709\\
\hline
0.474E-02&0.256E-03& 0.201& 0.151& 0.102&0.101& 165.0& 0.0 & 0.374\\
0.475E-02&0.256E-03& 0.195& 0.148& 0.101&0.110& 150.0& 0.0 & 0.370\\
0.475E-02&0.256E-03& 0.180& 0.141& 0.099&0.127& 135.0& 0.0 & 0.366\\
0.476E-02&0.255E-03& 0.171& 0.137& 0.097&0.140& 120.0& 0.0 & 0.364\\
0.476E-02&0.255E-03& 0.168& 0.135& 0.096&0.147& 105.0& 0.0 & 0.364\\
0.476E-02&0.255E-03& 0.166& 0.134& 0.096&0.150&  90.0&0.0 & 0.364\\
0.476E-02&0.255E-03& 0.168& 0.135& 0.096&0.147&  75.0& 0.0 &0.364\\
0.476E-02&0.255E-03& 0.171& 0.137& 0.097&0.140&  60.0&  0.0&0.365\\
0.476E-02&0.256E-03& 0.179& 0.141& 0.098&0.128&  45.0&  0.0& 0.367\\
0.475E-02&0.256E-03& 0.193& 0.148& 0.101&0.112&  30.0& 0.0 & 0.371\\
0.474E-02&0.256E-03& 0.198& 0.150& 0.101&0.104&  15.0& 0.0 & 0.375\\
0.475E-02&0.257E-03& 0.196& 0.149& 0.101&0.104&   0.0& 0.0 &0.379\\
\hline
0.474E-02&0.256E-03& 0.283& 0.141& 0.100&0.778& 165.0& 0.080&0.341\\
0.477E-02&0.256E-03& 0.213& 0.134& 0.095&0.115& 150.0& 0.042&0.333\\
0.477E-02&0.255E-03& 0.189& 0.123& 0.094&0.142& 135.0& 0.030&0.333\\
0.480E-02&0.255E-03& 0.166& 0.131& 0.088&0.155& 120.0& 0.025&0.332\\
0.478E-02&0.256E-03& 0.158& 0.126& 0.092&0.164& 105.0& 0.022&0.332\\
0.478E-02&0.255E-03& 0.159& 0.129& 0.092&0.161&  90.0& 0.022&0.332\\
0.479E-02&0.255E-03& 0.162& 0.118& 0.090&0.175&  75.0& 0.023&0.332\\
0.478E-02&0.256E-03& 0.164& 0.133& 0.092&0.151&  60.0& 0.025&0.331\\
0.477E-02&0.256E-03& 0.187& 0.121& 0.095&0.144&  45.0& 0.032&0.333\\
0.477E-02&0.256E-03& 0.221& 0.132& 0.095&0.113&  30.0& 0.045&0.333\\
0.475E-02&0.256E-03& 0.283& 0.147& 0.099&0.780&  15.0& 0.089&0.337\\
\hline
\end{tabular}
\end{center}
\end{table*}

%....................
\begin{table*}
\caption{Same that Table  \ref{tab:res-Bdb1} but with $\alpha^B=35, \ x_0^B=0.774,\ y_0^B=0.429, \ z_0^B=0.382  \ h_r=2.5 , \ h_z=0.158$.
The first set of results have been obtained with $x_1$ fixed to 0. In the second set of results $x_1$ is a free parameter to fit.
}
\label{tab:res-Bdb3}
\begin{center}
\begin{tabular}{ll|llllll|l}
\hline
Bulge   & Disk   &\multicolumn{6}{c|}{Bar}                     &                   \\
$\rho_0^B$  &$\rho_0^D$  &  $x_0^b$ & $y_0^b$ & $z_0^b$ &$\rho_0^b$ & $\alpha^b $ &  $x_1$ & $\chi^2$ \\
        &        & kpc    & kpc   &   kpc &         & deg        &  kpc &         \\
\hline
%0.618E-02&0.256E-03 & 1.073& 0.005& 0.028&0.159E+00&   0.0& 4.025\\
%0.619E-02&0.256E-03&  0.960& 0.006& 0.034&0.154E+00&  15.0& 3.950 \\
%0.613E-02&0.265E-03&  0.088& 0.036& 0.142&0.410E+00 & 30.0& 1.269\\
%0.637E-02&0.256E-03&  0.265& 0.060& 0.064&0.284E+00&  45.0& 0.506\\
%0.620E-02&0.256E-03&  0.269& 0.119& 0.098&0.107E+00&  60.0& 0.377\\
%0.621E-02&0.255E-03 & 0.260& 0.117& 0.097&0.300E-02&  75.0& 3.950\\
%0.621E-02&0.255E-03&  0.260& 0.117& 0.097&0.300E-02&  90.0& 3.939\\
%0.621E-02&0.255E-03&  0.260& 0.117& 0.097&0.300E-02& 105.0& 3.950\\
%0.620E-02&0.255E-03&  0.276& 0.117& 0.099&0.107E+00& 120.0& 0.374\\
%0.636E-02&0.256E-03&  0.223& 0.062& 0.066&0.278E+00& 135.0& 0.450\\
%0.640E-02&0.260E-03&  0.127& 0.027& 0.053&0.848E+00& 150.0& 0.727\\
%0.619E-02&0.656E-03 & 0.987& 0.004& 0.033&0.156E+00& 165.0& 4.008\\ 
%\hline
0.619E-02&0.256E-03& 0.205& 0.153& 0.102&0.100&  165.0& 0.0  & 0.382\\
0.619E-02&0.256E-03& 0.197& 0.150& 0.101&0.110&  150.0& 0.0  & 0.377\\
0.620E-02&0.256E-03& 0.182& 0.143& 0.099&0.126&  135.0& 0.0  & 0.374\\
0.621E-02&0.256E-03& 0.172& 0.138& 0.097&0.141&  120.0& 0.0  & 0.371\\
0.621E-02&0.255E-03& 0.169& 0.136& 0.097&0.147&  105.0& 0.0  & 0.371\\
0.621E-02&0.255E-03& 0.168& 0.136& 0.096&0.149&   90.0&0.0   & 0.371\\
0.621E-02&0.255E-03& 0.169& 0.136& 0.097&0.147&   75.0& 0.0  & 0.372\\
0.621E-02&0.256E-03& 0.172& 0.137& 0.097&0.141&   60.0&  0.0 & 0.373\\
0.620E-02&0.256E-03& 0.182& 0.142& 0.099&0.126&   45.0&  0.0 & 0.375\\
0.619E-02&0.256E-03& 0.196& 0.149& 0.101&0.110&   30.0& 0.0  & 0.379\\
0.619E-02&0.257E-03& 0.202& 0.151& 0.102&0.102&   15.0& 0.0  & 0.383\\
0.619E-02&0.257E-03& 0.201& 0.151& 0.102&0.101&    0.0& 0.0  & 0.386\\
\hline                                         
 0.605E-02&0.252E-03& 0.305& 0.144& 0.100&0.073& 165.0& 0.079&0.353\\
 0.608E-02&0.252E-03& 0.218& 0.135& 0.097&0.111& 150.0& 0.042&0.350\\
 0.608E-02&0.252E-03& 0.188& 0.126& 0.095&0.141& 135.0& 0.030&0.348\\
 0.609E-02&0.252E-03& 0.170& 0.126& 0.094&0.157& 120.0& 0.025&0.348\\
 0.610E-02&0.252E-03& 0.162& 0.122& 0.093&0.171& 105.0& 0.022&0.348\\
 0.610E-02&0.252E-03& 0.163& 0.119& 0.091&0.178&  90.0& 0.022&0.348\\
 0.611E-02&0.252E-03& 0.164& 0.124& 0.089&0.171&  75.0& 0.022&0.349\\
 0.611E-02&0.252E-03& 0.173& 0.127& 0.091&0.158&  60.0& 0.025&0.349\\
 0.609E-02&0.252E-03& 0.193& 0.126& 0.095&0.140&  45.0& 0.031&0.348\\
 0.609E-02&0.252E-03& 0.222& 0.137& 0.095&0.112&  30.0& 0.044&0.348\\
 0.606E-02&0.252E-03& 0.308& 0.149& 0.100&0.073&  15.0& 0.089&0.351\\
\hline
\end{tabular}
\end{center}
\end{table*}
%....................
\begin{table*}
\caption{Same that Table  \ref{tab:res-Bdb1} but with $\alpha^B=45, \ x_0^B=0.693,\ y_0^B=0.367, \ z_0^B=0.385  \ h_r=2.5 , \ h_z=0.159$. The first set of results have been obtained with $x_1$ fixed to 0. In the second set of results $x_1$ is a free parameter to fit.}
\label{tab:res-Bdb4}
\begin{center}
\begin{tabular}{ll|llllll|l}
\hline
Bulge   & Disk   &\multicolumn{6}{c|}{Bar}                     &                   \\
$\rho_0^B$  &$\rho_0^D$  &  $x_0^b$ & $y_0^b$ & $z_0^b$ &$\rho_0^b$ & $\alpha^b $ &  $x_1$ & $\chi^2$ \\
        &        & kpc    & kpc   &   kpc &         & deg     & kpc  &           \\
\hline
%0.763E-02&0.256E-03& 1.066&-0.002& 0.028&0.159E+00&  0.0& 4.135\\
%0.764E-02&0.256E-03& 0.992& 0.002& 0.031&0.157E+00& 15.0& 4.129\\
%0.789E-02&0.260E-03& 0.113& 0.025& 0.057&0.871E+00& 30.0& 0.836\\
%0.786E-02&0.256E-03& 0.242& 0.062& 0.064&0.277E+00& 45.0& 0.482\\
%0.766E-02&0.255E-03& 0.283& 0.119& 0.098&0.105E+00& 60.0& 0.385\\
%0.105E-01&0.396E-03& 1.076& 0.191& 0.139&0.151E-01& 75.0& 11.99 \\
%0.104E-02&0.245E-03& 2.961& 0.365& 0.239&0.478E-02& 90.0&   7.728\\
%0.631E-02&0.272E-03& 1.927& 0.285& 0.193&0.111E-01&105.0&   3.662\\
%0.766E-02&0.255E-03& 0.294& 0.119& 0.099&0.103E+00&120.0&  0.385\\
%0.786E-02&0.256E-03& 0.247& 0.063& 0.065&0.273E+00&135.0& 0.476\\
%0.790E-02&0.259E-03& 0.175& 0.027& 0.048&0.857E+00&150.0& 0.698\\
%0.764E-02&0.256E-03& 0.994& 0.008& 0.033&0.156E+00&165.0& 3.916\\
%\hline
0.761E-02&0.255E-03& 0.211& 0.167& 0.105&0.095& 165.0&  0.0 &0.389\\
0.763E-02&0.255E-03& 0.204& 0.154& 0.103&0.106& 150.0&  0.0 &0.382\\
0.766E-02&0.255E-03& 0.184& 0.143& 0.099&0.126& 135.0&  0.0 &0.378\\
0.767E-02&0.256E-03& 0.173& 0.138& 0.097&0.141& 120.0&  0.0 &0.376\\
0.767E-02&0.255E-03& 0.170& 0.136& 0.097&0.147& 105.0&  0.0 &0.375\\
0.767E-02&0.255E-03& 0.168& 0.136& 0.096&0.149&  90.0& 0.0  & 0.376\\
0.767E-02&0.255E-03& 0.170& 0.136& 0.097&0.147&  75.0&  0.0 &0.376\\
0.767E-02&0.256E-03& 0.172& 0.138& 0.097&0.141&  60.0&   0.0&0.377\\
0.766E-02&0.255E-03& 0.183& 0.143& 0.099&0.126&  45.0&   0.0&0.379\\
0.763E-02&0.256E-03& 0.204& 0.151& 0.102&0.107&  30.0&  0.0 &0.384\\
0.761E-02&0.256E-03& 0.217& 0.161& 0.106&0.092&  15.0&  0.0 &0.391\\
0.764E-02&0.257E-03& 0.204& 0.153& 0.102&0.099&   0.0&  0.0 &0.390\\
\hline
0.766E-02&0.255E-03& 0.288& 0.150& 0.098&0.076& 165.0& 0.079&0.350\\
0.767E-02&0.255E-03& 0.220& 0.134& 0.096&0.112& 150.0& 0.042&0.347\\
0.768E-02&0.255E-03& 0.186& 0.126& 0.096&0.141& 135.0& 0.030&0.347\\
0.770E-02&0.255E-03& 0.166& 0.134& 0.092&0.151& 120.0& 0.025&0.345\\
0.771E-02&0.255E-03& 0.163& 0.122& 0.090&0.171& 105.0& 0.022&0.346\\
0.771E-02&0.255E-03& 0.160& 0.119& 0.090&0.178&  90.0& 0.022&0.346\\
0.771E-02&0.255E-03& 0.165& 0.122& 0.090&0.170&  75.0& 0.022&0.346\\
0.770E-02&0.255E-03& 0.171& 0.122& 0.092&0.162&  60.0& 0.025&0.346\\
0.769E-02&0.255E-03& 0.193& 0.126& 0.094&0.140&  45.0& 0.031&0.346\\
0.768E-02&0.255E-03& 0.227& 0.133& 0.095&0.112&  30.0& 0.045&0.347\\
0.765E-02&0.256E-03& 0.308& 0.146& 0.100&0.073&  15.0& 0.090&0.350\\
\hline
\end{tabular}
\end{center}
\end{table*}
\bibliographystyle{aa}
\bibliography{/home/rodrigue/biblio_bibtex/bib_milky-way,/home/rodrigue/biblio_bibtex/bib_general,/home/rodrigue/biblio_bibtex/bib_ISM,/home/rodrigue/biblio_bibtex/bib_GC,/home/rodrigue/biblio_bibtex/bib_galaxies}
\end{document}